\documentclass[11pt,prd,a4paper,preprintnumbers,amsmath,amssymb,nofootinbib]{article}
\pdfoutput=1
\usepackage{feynmp-auto,expdlist}
\usepackage{amsmath,amsfonts,amssymb}
\usepackage{graphicx}
\usepackage{enumerate}
\usepackage{hyperref}
\usepackage{latexsym}
\usepackage{hepnicenames}
\usepackage{enumerate}
\usepackage{soul}
\usepackage[normalem]{ulem}
\usepackage{wasysym}
\usepackage{makecell}
\usepackage{bbm}
\usepackage{physics}

\font\tenrsfs=rsfs10 at 12pt
\font\sevenrsfs=rsfs7
\font\fiversfs=rsfs5
\newfam\rsfsfam
\textfont\rsfsfam=\tenrsfs
\scriptfont\rsfsfam=\sevenrsfs
\scriptscriptfont\rsfsfam=\fiversfs

\numberwithin{equation}{section}

\usepackage{mathrsfs}
\usepackage{braket}
\usepackage{titling}
\usepackage{amsmath}
\usepackage{slashed}
\usepackage{amssymb}
\usepackage{epsfig}
\usepackage{graphicx}
\usepackage{color}
\usepackage{rotating}
\usepackage{hyperref}
\usepackage[margin=1.in]{geometry}
\usepackage[table,xcdraw,dvipsnames]{xcolor}
\usepackage[compress,numbers,sort]{natbib}
\usepackage{colortbl}
\usepackage{pdflscape}
\usepackage{color}
\usepackage{mathtools}
\usepackage{colortbl}
\usepackage{comment}

\newcommand{\U}{{\rm U}}

\newcommand{\F}{{\cal F}}

\newcommand{\M}{{\cal M}}
\renewcommand{\L}{{\cal L}}
\newcommand{\N}{{\cal N}}
\newcommand{\E}{{\cal E}}
\newcommand{\Y}{{\cal Y}}

\definecolor{nicered}{rgb}{0.7,0.1,0.1}
\definecolor{nicegreen}{rgb}{0.1,0.5,0.1}
\definecolor{red}{rgb}{1.0, 0, 0}
\definecolor{niceblue}{rgb}{0,0,0.8}
\allowdisplaybreaks

\definecolor{blus}{cmyk}{1,1,0,0.6}
\definecolor{verde}{cmyk}{0.92,0,0.59,0.25}
\definecolor{rossos}{cmyk}{0,1,1,0.55}

\hypersetup{colorlinks,bookmarksopen,bookmarksnumbered,linkcolor=blus,pdfstartview=FitH,urlcolor=rossos,citecolor=verde}

\def\eq#1{{Eq.~(\ref{#1})}}
\def\eqs#1#2{{Eqs.~(\ref{#1})--(\ref{#2})}}
\def\fig#1{{Fig.~\ref{#1}}}

\renewcommand{\bar}{\overline}

\newcommand{\beq}{\begin{equation}}
\newcommand{\eeq}{\end{equation}}
\newcommand{\bea}{\begin{eqnarray}}
\newcommand{\eea}{\end{eqnarray}}

\renewcommand{\[}{\left[}
\renewcommand{\]}{\right]}
\renewcommand{\(}{\left(}
\renewcommand{\)}{\right)}

\def\be{\begin{equation}}
\def\ee{\end{equation}}

\begin{document}

\begin{center}  
{\huge
\bf\color{blus} 
Closing in on new chiral leptons at the LHC} \\
\vspace{0.8cm}

{\bf Daniele Barducci$^{a}$, Luca Di Luzio$^{b}$, Marco Nardecchia$^{c}$, Claudio Toni$^{c,d,b}$ }\\[7mm]

{\it $^a$Dipartimento di Fisica ``Enrico Fermi'', Universit\`a di Pisa and INFN, Sezione di Pisa, \\ Largo Bruno Pontecorvo 3, I-56127 Pisa, Italy}\\[1mm]
{\it $^b$Istituto Nazionale di Fisica Nucleare (INFN), Sezione di Padova, \\
Via F. Marzolo 8, 35131 Padova, Italy}\\[1mm]
{\it $^c$Physics Department and INFN Sezione di Roma La Sapienza, \\ 
Piazzale Aldo Moro 5, 00185 Roma, Italy}\\[1mm]
{\it $^d$Dipartimento di Fisica e Astronomia `G.~Galilei', Universit\`a di Padova,
 \\ Via F. Marzolo 8, 35131 Padova, Italy
}

\vspace{0.3cm}
\begin{quote}
 
We study the phenomenological viability of 
chiral
extensions of the Standard Model, 
with new chiral fermions acquiring 
their mass 
through interactions with a single Higgs.
We examine constraints from 
electroweak precision tests, 
Higgs physics and direct searches at the LHC.  
Our analysis indicates 
that purely chiral scenarios are  
perturbatively 
excluded by the combination of Higgs coupling measurements 
and LHC direct searches. 
However, allowing for 
a partial contribution from vector-like masses 
opens up the parameter space and non-decoupled 
exotic leptons 
could account for the observed 2$\sigma$ 
deviation 
in $h \to Z\gamma$. This scenario will be 
further tested in the  
high-luminosity phase 
of the LHC.


\end{quote}

\thispagestyle{empty}

\end{center}

\bigskip
\setcounter{tocdepth}{2}
\tableofcontents

\clearpage

\section{Introduction}
\label{sec:intro}

Before the discovery of the Higgs boson and the accurate determination of its couplings to gauge bosons and third-generation fermions, 
the possibility of a fourth generation of Standard Model (SM) fermions was a very popular scenario, see {\it e.g.}~\cite{Holdom:2009rf}. However, 
due to its non-decoupling 
nature, purely chiral extensions of the SM imply 
sizeable modifications in 
loop-induced Higgs couplings to gluons and photons, 
so that a fourth generation of SM fermions is nowadays 
ruled out, at least perturbatively \cite{Kribs:2007nz,Kuflik:2012ai,Eberhardt:2012gv}. 
 
A comprehensive classification of chiral extensions of the SM 
beyond fourth generation  
was given in Ref.~\cite{Bizot:2015zaa}, according to the following requirements: 
$i)$ no massless fermions after electroweak symmetry breaking (EWSB); 
$ii)$ no SM gauge anomalies; 
$iii)$ non-zero corrections to the Higgs boson couplings. 
Phenomenological viability basically requires the new chiral fermions to be uncolored. In fact, LHC direct searches push the mass of colored states above the TeV and, due to their non-decoupling nature, they in turn modify 
by $\mathcal{O}(1)$ value the gluon-gluon fusion amplitude for the Higgs production, in stark contradiction with Higgs data. 
Focussing on uncolored states, hereafter named as exotic leptons, 
Ref.~\cite{Bizot:2015zaa} identified a minimal set 
of chiral leptons, {\it cfr.}~\eq{eq:fermions_QN}, whose mass 
stems completely from EWSB and which at that time was still 
perturbatively viable 
when considering a combination of phenomenological constraints 
from electroweak precision tests, 
Higgs physics and direct searches at the LHC.

Incidentally, the same chiral field content above was invoked  
in specific ultraviolet (UV) completions of some beyond the SM (BSM) scenarios. In Ref.~\cite{Bonnefoy:2020gyh} 
it was discussed in the context of an axion model 
featuring axion couplings to electroweak gauge bosons, 
which were generated after integrating out the heavy chiral leptons charged under a global $U(1)$ symmetry. Another scenario pertains instead the UV completion of a vector boson  
associated to a $U(1)_{X}$ gauge symmetry, with $X$ an anomalous combination of the SM global symmetries {\it e.g.}~$B+L$. 
In this case, the exotic chiral leptons, also charged under $U(1)_X$, are also known as {\it anomalons}, 
{\it i.e.}~the fermion sector responsible for the cancellation 
of gauge anomalies involving $U(1)_{X}$ and the electroweak 
gauge factors \cite{Duerr:2013dza,Duerr:2013lka,Dobrescu:2014fca,Dobrescu:2015asa}. The role of the anomalons is especially relevant 
in the case of the effective field theory of a light 
vector boson coupled to an anomalous SM current
since, as argued in Refs.~\cite{Dror:2017ehi,Dror:2017nsg},
the vector boson interactions feature an axion-like behaviour 
leading to processes 
enhanced as $(\text{energy} / m_X)^2$, thus resulting into the typically most stringent bounds on the vector boson parameter space
stemming, {\it e.g.},~from flavour-violating process. 
In Ref.~\cite{DiLuzio:2022ziu} it was shown that this conclusion relies on the UV completion of the theory
and, in particular, the bounds of Refs.~\cite{Dror:2017ehi,Dror:2017nsg} are completely evaded 
if the anomalon fields pick up their mass entirely from the 
Higgs, that is precisely the phenomenological scenario considered in Ref.~\cite{Bizot:2015zaa}. 

In this work we revisit the phenomenological analysis of Ref.~\cite{Bizot:2015zaa}, mainly in view of the updated 
limits from direct searches at the LHC, which have dramatically improved in the last decade. In particular, there are two types of searches which need to be considered, depending on whether the exotic leptons mix or not with the SM leptons. 
In the former case, they are unstable and strong limits can be obtained by recasting same-sign lepton searches, while in the latter case they are stable, leading to striking signatures at collider, in terms of charged tracks. 
A further reason to reconsider the scenarios above is 
given by the recent ATLAS and CMS combination for the search 
of the $h \to Z\gamma$ decay channel, whose signal strength is found to be $R_{Z\gamma} =2.2\pm 0.7$~\cite{CMS:2023mku}, 
thus exhibiting a mild $\sim 2\sigma$ tension with respect to the SM. 

The paper is structured as follows. In Sec.~\ref{sec:} 
we recall the classification of SM chiral extensions of \cite{Bizot:2015zaa}, also in light of the new Higgs data. 
Sec.~\ref{sec:pheno} deals with the phenomenology of 
purely chiral extensions of the SM, while in Sec.~\ref{sec:pheno_VL} we include the contribution 
of vector-like components to the masses of the 
exotic leptons. This allows to open up the parameter space of the model, providing a generalized setup in order to interpret the 
new $h \to Z\gamma$ measurement. We finally conclude in 
Sec.~\ref{sec:concl}, while more technical details on the classification of chiral SM extensions and general formulae for 
loop observables are deferred to App.~\ref{app1} an App.~\ref{app2}, 
respectively. 

\section{Chiral extensions of the Standard Model}
\label{sec:}

Purely chiral extensions of the SM are strongly constrained 
by Higgs couplings measurements. In particular via the loop-induced couplings to gluons and photons, which feature a non-decoupling contribution from heavy chiral fermions acquiring their mass entirely from the Higgs vacuum expectation value (VEV). 
In this work we assume the presence of a single Higgs doublet, from which all the chiral fermions 
acquire their mass. 
However, 
extended scalar sectors can enlarge the parameter 
space of Higgs couplings and allow for 
new chiral fermions with SM-like couplings 
(see 
e.g.~\cite{Banerjee:2013hxa,Holdom:2014bla,Das:2017mnu}).
Given that gluon fusion stands as the predominant mechanism for Higgs production, the introduction of new colored chiral fermions would significantly spoil various Higgs coupling measurements \cite{Kribs:2007nz,Kuflik:2012ai,Eberhardt:2012gv}. 
Furthermore, new colored fermions are subject to stringent constraints from direct searches at the LHC which require their mass to be larger than $\sim1\;$TeV, thus pushing the associated Yukawa couplings with the Higgs beyond the perturbativity regime. Hence, in the following we will focus on 
color-singlet chiral extensions of the SM with non-trivial 
electroweak quantum numbers. 

The logic for such classification is to find an anomaly-free set of new chiral fermions, whose BSM contribution to $h\to\gamma\gamma$ interferes with the SM one such that the decay rate is again SM-like.
While this classification 
was discussed in Ref.~\cite{Bizot:2015zaa} (see also \cite{DAgnolo:2023rnh}), 
we revisit it in App.~\ref{app1} also in view of the recent hint 
for a $h \to Z \gamma$ signal above the SM value, 
following the 
ATLAS and CMS combination which recently measured $R_{Z\gamma} =2.2\pm 0.7$~\cite{CMS:2023mku}. 
The bottom-line is that the minimal set of new chiral leptons allowing for a SM-like Higgs coupling to photons 
is given by \cite{Bizot:2015zaa}
\be\label{eq:fermions_QN}
\L_{L,R} = 
\begin{pmatrix}
\N_\L \\
\E_\L 
\end{pmatrix}_{L,R}
\sim ({\bf{1}},{\bf{2}})_Y \ , \quad 
\E_{L,R} \sim ({\bf{1}},{\bf{1}})_{Y-\frac{1}{2}} \ , \quad 
\N_{L,R} \sim ({\bf{1}},{\bf{1}})_{Y+\frac{1}{2}} \ , 
\ee
which come in opposite chiralities, $L$ and $R$,
in order to cancel 
SM gauge anomalies.\footnote{Note that the field content in \eq{eq:fermions_QN} 
is formally vector-like under the SM gauge group. 
In this paper we use the term {\it chiral} fermion 
to indicate the absence of a vector-like mass term, 
which could be forbidden by an extra symmetry beyond 
that of the SM, {\it cfr.}~discussion below \eq{eq:VLmass}.} 
Note that for $Y=-\frac{1}{2}$ the exotic leptons have respectively the same quantum numbers of the SM left-handed doublet $L$, the right-handed singlet $e$ and a would be right-handed neutrino.

The field content in \eq{eq:fermions_QN} allows for  the following Yukawa interactions with the SM Higgs doublet 
\be
\label{eq:yukawa}
-\L_y = 
y_1 \bar \L_L H \E_R + 
y_2 \bar \L_R H \E_L +
y_3 \bar\L_L \tilde H \N_R + 
y_4 \bar\L_R \tilde H \N_L + 
 h.c. \ ,
\ee
where $\tilde H = i \sigma_2 H^*$ and $\sigma_2$ is the second Pauli matrix. 
In principle, the SM symmetry would also permit vector-like masses for the new fermions,
\be
\label{eq:VLmass}
-\Delta\L_y = 
m_\L \bar \L_L  \L_R + 
m_\N \bar \N_L  \N_R + 
m_\E \bar \E_L  \E_R +
 h.c. \ , 
\ee
which can be forbidden by means of a discrete symmetry or via 
an extra $U(1)_X$ gauge symmetry that acts differently 
on the $L$ and $R$ components of the exotic leptons. 
This latter possibility actually arises in 
the UV complete models discussed in Ref.~\cite{DiLuzio:2022ziu}, 
where $X$ is an anomalous symmetry of the SM, {\it e.g.}~$B+L$, 
and the fields in \eq{eq:fermions_QN} are introduced in order to cancel the gauge anomalies between $U(1)_X$ and the SM gauge group.

Moreover, for specific values of $Y$ additional Yukawa interactions are possible, 
allowing for a mixing of the exotic leptons with the SM leptons. 
Those terms are phenomenologically important and will be discussed in Sec.~\ref{sec:prompt}. 


After EWSB the interactions in Eq.~\eqref{eq:yukawa} and Eq.~\eqref{eq:VLmass} produce a spectrum with two 2-flavor Dirac fermions  $\Psi^{\E} = (\E_\L, \E)$ and  $\Psi^\N = (\N_\L, \N)$, with electric charge $Q = Y\mp \frac{1}{2}$ respectively. The mass Lagrangian 
\be
-\L_{\rm mass} = 
\bar \Psi^\E_L {\cal M}_\E   \Psi^\E_R +
\bar \Psi^\N_L {\cal M}_\N   \Psi^\N_R + h.c. \ ,
\ee
with mass matrices
\be
{\cal M}_\E =
\begin{pmatrix}
m_\L & \frac{v}{\sqrt{2}}y_1 \\
\frac{v}{\sqrt{2}}y_2^* & m_\E \\
\end{pmatrix} \ ,
\qquad
{\cal M}_\N =
\begin{pmatrix}
m_\L & \frac{v}{\sqrt{2}}y_3 \\
\frac{v}{\sqrt{2}}y_4^* & m_\N \\
\end{pmatrix} \, , 
\ee
can be diagonalized by two bi-unitary transformations, yielding the eigenstates 
$\Psi^{\E_{1,2}}$ and $\Psi^{\N_{1,2}}$
with masses $m_{\Psi^{\E_{1,2}}}$ and $m_{\Psi^{\N_{1,2}}}$ respectively.
Note that in the limit $m_\N=m_\E$, $y_1=y_3$ and $y_2=y_4$, hence $\M_\N=\M_\E$, the
Lagrangian sector $\L_y+\Delta\L_y$ features a custodial symmetry which helps in taming corrections to electroweak
precision observables.

In the following, we will first discuss in Sec.~\ref{sec:pheno} the phenomenology of purely chiral fermions, {\it i.e.}~the limit of vanishing vector-like masses and then, in Sec.~\ref{sec:pheno_VL}, we will analyze the impact of vector-like masses on the parameter space of the model, especially in view of the recent hint of a mild excess in the $h \to Z\gamma$ decay rate.

\section{Phenomenology of new chiral leptons}
\label{sec:pheno}

Among the most relevant phenomenologival observables 
to constrain the parameter space of new chiral leptons, 
in this section 
we consider in turn electroweak precision tests, loop-induced Higgs couplings and direct searches, as well as theoretical limits imposed 
by perturbative unitarity. 

\subsection{Electroweak precision tests}

The new chiral fermions contribute to the Peskin-Takeuchi $S$ and $T$ parameters~\cite{Peskin:1990zt,Peskin:1991sw} as~\cite{Bizot:2015zaa}
\be
\begin{split}
S & = \frac{1}{6\pi} 
\[ \(1 - 2 Y \log\frac{m^2_{\Psi^{\N_1}}}{m^2_{\Psi^{\E_1}}}\) 
+ \(1 + 2 Y \log\frac{m^2_{\Psi^{\N_2}}}{m^2_{\Psi^{\E_2}}}\) 
+ \mathcal{O}\(\frac{m^2_{Z}}{m^2_{\Psi^{\N,\, \E}}}\)
\] \ , \\
T & = \frac{1}{16\pi c^2_w s^2_w m^2_Z} 
\( m^2_{\Psi^{\N_1}} + m^2_{\Psi^{\E_1}} 
- 2 \frac{m^2_{\Psi^{\N_1}}m^2_{\Psi^{\E_1}}}{m^2_{\Psi^{\N_1}}-m^2_{\Psi^{\E_1}}} \log\frac{m^2_{\Psi^{\N_1}}}{m^2_{\Psi^{\E_1}}}\) \\
&+ \frac{1}{16\pi c^2_w s^2_w m^2_Z} 
\( m^2_{\Psi^{\N_2}} + m^2_{\Psi^{\E_2}} 
- 2 \frac{m^2_{\Psi^{\N_2}}m^2_{\Psi^{\E_2}}}{m^2_{\Psi^{\N_2}}-m^2_{\Psi^{\E_2}}} \log\frac{m^2_{\Psi^{\N_2}}}{m^2_{\Psi^{\E_2}}}\)  \ .\\
\end{split}
\ee
The latest fits for the oblique parameters give \cite{Gfitter}:
$S = 0.05 \pm 0.11$ and 
$T =0.09 \pm 0.13$, 
which are easily satisfied in the custodial limit $y_1 = y_3$ and $y_2 = y_4$, where $m_{\Psi^{\E_1}}=m_{\Psi^{\N_1}}$ and $m_{\Psi^{\E_2}}=m_{\Psi^{\N_2}}$.
Note, also, that contributions beyond $S$ and $T$, stemming 
from the loop-induced electroweak 
corrections to neutral and charged Drell-Yan processes,  
yield bounds that 
are weaker than current direct searches 
(to be discussed in 
Sec.~\ref{sec:direct}), even when considering 
projections at the high-luminosity phase of LHC (see {\it e.g.}~Ref.~\cite{DiLuzio:2018jwd}).

\subsection{Higgs physics}
\label{sec:higgs}

As well known, new fermions which acquire mass from Yukawa interactions with the Higgs field contribute to
various Higgs observables. In particular, they modify the predictions for the $h\to \gamma\gamma$ and $h\to Z\gamma$ signal strengths, $R_{\gamma\gamma}$ and $R_{Z\gamma}$. These observables are defined as the ratio of the cross-section into the corresponding final state with respect to the SM ones. In the case under consideration of uncolored new states, they can be casted as a ratio of the BSM over the SM branching ratios, which can in turn be expressed in terms of the corresponding squared amplitudes. In particular, by assuming that the new fermions are much heavier than the Higgs boson and with the field content of Eq.~\eqref{eq:fermions_QN}, one has
\be
R_{\gamma \gamma, Z\gamma } = \frac{|A_{\gamma\gamma,Z\gamma}^{\rm SM}+A_{\gamma\gamma,Z\gamma}^{\rm BSM}|^2}{|A_{\gamma\gamma,Z\gamma}^{\rm SM}|^2} \, ,
\ee
with $A_{\gamma\gamma}^{\rm SM} \simeq -6.5  $, $A_{Z\gamma}^{\rm SM} \simeq -6.64$ and 
\be
\label{A1}
\begin{split}
& A_{\gamma\gamma}^{\rm BSM} \simeq  \frac{4}{3}\left(1+  4 Y^2 \right) \ ,\\
& A_{Z\gamma}^{\rm BSM} \simeq  \frac{2}{3}\left[1 -   (1+8Y^2) {\rm tg}_w^2 \right] \  ,
\end{split}
\ee
where ${\rm tg}_w$ is the tangent of the Weinberg angle. The most recent measurement of the $h\to \gamma\gamma$ signal strength is $R_{\gamma \gamma} =1.00\pm 0.12$~\cite{ATLAS:2019nkf}. As regarding $h\to Z\gamma$, the ATLAS and CMS collaboration have recently released a combined measurement which represent the first evidence for this radiative Higgs decay mode. The signal strength is measured to be  $R_{Z\gamma} =2.2\pm 0.7$~\cite{CMS:2023mku}. While the $\gamma\gamma$ mode appear to be in perfect agreement with the SM predictions, the $Z\gamma$ one exhibits a mild $\sim 2\sigma$ tensions. 

\begin{figure}[t!]
\centering
\includegraphics[width=0.48\textwidth]{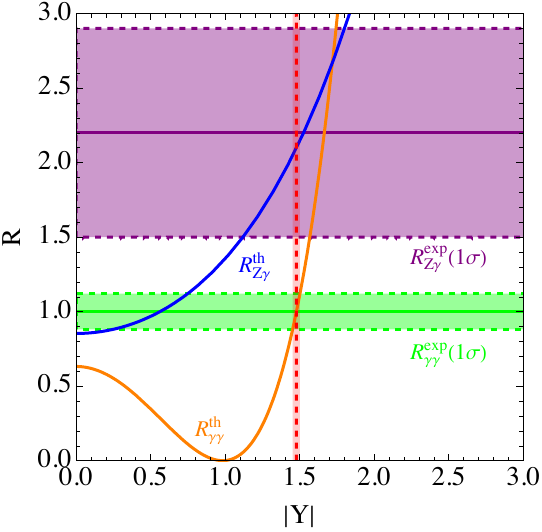} 
\caption{$R_{\gamma\gamma}$ and $R_{Z\gamma}$ signal strengths as a function of the hypercharge $Y$ of the fermion $\L$. The area between the shaded lines around the central value corresponds to the current $1\sigma$ experimental uncertainty.}
\label{fig:R_az}
\end{figure}

The predictions of the model with the field content of Eq.~\eqref{eq:fermions_QN} for these signal strength are shown in Fig.~\ref{fig:R_az} in function of the value $Y$ of the hypercharge for the BSM fermion $\L$, where the shaded areas around the central values represent the $1\sigma$ uncertainty bands. Remarkably, by imposing  the model to reproduce the experimental central value for $R_{\gamma\gamma}=1.00$, one obtains the condition $Y\sim \pm1.48$, which leads to the prediction $R_{Z\gamma}\sim 2.10$, in notable agreement with the ATLAS and CMS combined measurement~\cite{CMS:2023mku}. It is also interesting to note that the obtained hypercharge value is close to the semi-integer values $Y=\pm\frac{3}{2}$, which are more theoretically appealing.

Given these constraints, in the following we will thus analyse the phenomenological consequence of the model by fixing $Y\simeq -\frac{3}{2}$, while the case $Y\simeq +\frac{3}{2}$ is simply obtained by charge conjugation.

\subsection{Direct searches}
\label{sec:direct}

\begin{figure}[t!]
\centering
\includegraphics[width=0.48\textwidth]{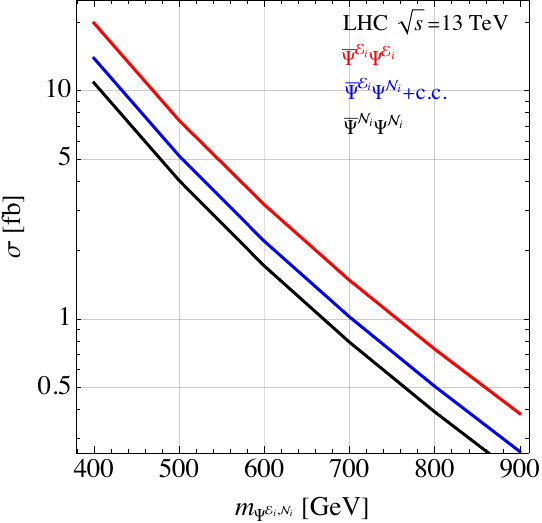} 
\caption{Cross-section for pair-production of a pair of exotic fermions at the 13\;TeV LHC. In the case of CC interactions we assume degenerate states
$m_{\Psi^{\E_i}}=m_{\Psi^{\N_i}}$.
}
\label{fig:pair_prod}
\end{figure}

With the choice $Y\simeq - \frac{3}{2}$ for the hypercharge  of the doublet $\L$, the two 2-flavor Dirac fermions $\Psi^\E$ and $\Psi^\N$ have electric charge $Q\simeq -2e$ and $Q\simeq -e$ respectively. At the LHC they can be produced in pair via charged-current (CC) and neutral-current (NC) Drell-Yan 
(DY) interactions. In the mass eigenbasis their interactions read
\be
\begin{split}
& \L_{\rm CC} = \frac{g}{\sqrt 2}
\bar \Psi^{\N_{1}} \gamma^\mu P_L \Psi^{\E_{1}} W_\mu^+ +
\frac{g}{\sqrt 2}
\bar \Psi^{\N_{2}} \gamma^\mu P_R \Psi^{\E_{2}} W_\mu^+ +
h.c. \ , \\
& \L_{\rm NC} = \frac{g}{c_w} \bar \Psi^{\E_i} \gamma^\mu (g^{\E_i}_L P_L +g^{\E_i}_R P_R) \Psi^{\E_i} Z_\mu +
 \frac{g}{c_w} \bar \Psi^{\N_i} \gamma^\mu (g^{\N_i}_L P_L +g^{\N_i}_R P_R) \Psi^{\N_i} Z_\mu \ ,
 \end{split}
\ee
with $P_{L,R} = (1\mp \gamma_5)/2$ and with 
\be
\begin{split}
& g^{\E_1}_L  = g^{\E_2}_R = -\frac{1}{2}+ 2 s_w^2  \ , \qquad  g^{\E_1}_R  = g^{\E_2}_L =  2 s_w^2 \ ,   \\
& g^{\N_1}_L  = g^{\N_2}_R =+ \frac{1}{2}+  s_w^2  \ , \qquad  g^{\N_1}_R  = g^{\N_2}_L =   s_w^2    \ .
\end{split}
\ee
We show in Fig.~\ref{fig:pair_prod} the cross-sections for the production of a pair of new fermions in NC and CC interactions, where in the latter case we assume the custodial limit and the fermions are mass degenerate. The largest rates are the ones for the production of doubly-charged states. As regarding their decay modes, these depends on whether the exotic fermions mix or not with SM fields. We discuss these two scenarios separately.

\subsubsection{$Y\ne -\frac{3}{2}$: no mixing with SM leptons}
\label{sec:HSCP}

If $Y\ne -\frac{3}{2}$, no mixing terms are allowed. In this case for each  flavor $i$ the heaviest between the $\Psi^{\E_i}$ and $\Psi^{\N_i}$ states decays via CC interactions into the lightest one, which is stable because of exotic lepton number conservation. Charged stable states are cosmologically dangerous and largely excluded unless they are not produced in the early Universe or some mechanism dilutes their abundance. They also provide striking signatures at colliders, in terms of charged tracks.  

The strongest limit on stable states with $Q=2e$ is set by the ATLAS analysis~\cite{ATLAS:2023zxo} which searches for stable particles with various choices for their electric charge, ranging from $Q=2e$ to $Q=7e$, and specific coupling structure to the $Z-$boson. They consider production via both DY and photon-fusion interactions at 
$\sqrt s=13\;$TeV with 139\;fb$^{-1}$ of integrated luminosity.
By properly taking into account the different coupling structure of the model we are investigating with respect to the one assumed by the ATLAS analysis (see e.g.~\cite{DiLuzio:2015oha}), one obtains a limit of $m_{\Psi^{{\E}_i}}\gtrsim 1030\;$GeV for a single flavor of doubly charged exotic leptons. This limit increases to $\sim 1140\;$GeV for two degenerate doubly charged exotic leptons. 

This analysis does not consider the case of singly charged heavy leptons with $Q=e$. In the scenario where this state is the lightest present in the spectrum, the strongest limit is set by the CMS analysis~\cite{CMS:2016kce}, performed at $\sqrt s=13\;$TeV with 3.2\;fb$^{-1}$ of integrated luminosity. By recasting the CMS results one obtains a limit $m_{\Psi^{{\N}_i}}\gtrsim 540\;$GeV, with the available data set for a single flavor of singly charged exotic lepton, limits that increases to $\sim 630\;$GeV, for two degenerate singly charged exotic leptons. By projecting the results of this analysis to an integrated luminosity of 139\;fb$^{-1}$, corresponding to the luminosity of the ATLAS search for doubly charged leptons, these limits increase to 820\;GeV and 930\;GeV respectively.

\subsubsection{$Y=-\frac{3}{2}$: exotic fermions mixed with SM leptons}
\label{sec:prompt}

If $Y= -\frac{3}{2}$ additional Yukawa interactions among the exotic and SM leptons are allowed. In particular one has
\be
-{\L}_{\rm mix} = \lambda_{i,R} \bar L_L^i H \N_R + \lambda_{i,L} \bar\L_L \tilde H e_R^i  + h.c. \ ,
\ee
which modifies the mass matrix in the $Q=e$ sector. After EWSB we obtain
\be
- {\cal L}_{\rm mix} = \frac{v}{\sqrt 2}
\begin{pmatrix}
\bar e^i \\
\bar \N_{\cal L} \\
\bar \N 
\end{pmatrix}_L
\begin{pmatrix}
y^{ij}  &~~\lambda_{i,R} &~~0 \\
\lambda^T_{j,L}&~~y_3 &~~ 0  \\
0  & 0&~~y_4^* \\
\end{pmatrix}
\begin{pmatrix}
e^j \\
\N \\
\N_{\cal L}
\end{pmatrix}_R + h.c. \ ,
\ee
which shows that only one of the exotic $Q=e$ fermions mixes with the SM leptons.
Importantly, while the heaviest among $\Psi^{\E_i}$ and $\Psi^{\N_i}$ always decays into the lightest one via CC interactions, 
this mixing also triggers the decay of the lightest exotic fermions into SM leptons, thus producing a radically different phenomenology with respect to the one of stable charged particles discussed in Sec.~\ref{sec:HSCP}.
Allowing for the presence of a term like ${\cal M}_{\rm mix}^{j} \bar\N_L e_R^j + h.c.$, produces a mixing among all the $Q=e$ eigenstates, so that for both flavors the lightest state is always unstable.
The mixing couplings among anomalons and SM leptons modify the $Z$ couplings to SM charged leptons, which are very well measured, and potentially induce flavour changing violating decays $Z\to e\mu,\mu\tau,\tau e$. Both the constraints require the mixing angles to be roughly $\lesssim\mathcal{O}(10^{-3})$ and hence their insertion does not modify the calculation of the Higgs signal strength in \eq{A1}.  
In a similar way, the CC interactions become 
\be
\L_{\rm CC} = \frac{g}{\sqrt 2}\alpha_i
\bar \Psi^{\E_{1}} \gamma^\mu P_L e^i W_\mu^- +
\frac{g}{\sqrt 2}\beta_i
\bar \Psi^{\E_{2}} \gamma^\mu P_R e^i W_\mu^- +
h.c. \ ,
\ee
where $\alpha_i$ and $\beta_i$ are mixing angles whose explicit expressions and magnitude are not relevant for the rest of the discussion as long as they are $\gtrsim \mathcal{O}(10^{-7})$, so that doubly charged states promptly
decay into a final state with a same-sign (SS) lepton pair via
\be
\Psi^{\E_i} \to W^- \ell^- \to \ell^- \ell^- \slashed E_T \ .
\ee
Final states with SS leptons have small SM backgrounds. Together with the fact that the 
DY cross-section for the production of a $\Psi^{\E_i} \bar \Psi^{\E_i}$ pair is the largest, see Fig.~\ref{fig:pair_prod}, we expect this process to yield the strongest limit in the scenario where the exotic fermions mix with the SM leptons. 

No direct searches for doubly charged leptons exist. There exists however an ATLAS analysis targeting pair-produced doubly charged scalars decaying into a SS lepton pair, performed at $\sqrt s=13\;$TeV with 36.1\;fb$^{-1}$ of integrated luminosity~\cite{ATLAS:2017xqs}. In order to recast this analysis we  have implemented the model into the {\tt Feynrules}~\cite{Alloul:2013bka} package using the {\tt UFO}~\cite{Degrande:2011ua} format in order to perform a simulation with {\tt MadGraph}~\cite{Alwall:2014hca}.
The ATLAS search defines 8 mutually exclusive signal regions (SRs), each of them featuring at least one SS lepton pair, categorizing them with respect to lepton multiplicity and flavor.

The basic selection cuts imposed are $p_T^\ell > 30\;$GeV, $|\eta_e|<2.37$ excluding the crack region $1.37 < |\eta_e| < 1.52$ and $|\eta_\mu |<2.5$. Given that the ATLAS analysis is a simple leptonic cut-and-count search, we decide to work at the parton level, neglecting parton showering and detector reconstruction effects. We expect that the results that we obtain are a good approximation of a more sophisticated procedure. In order to mimic a jet clustering algorithm, needed to apply the lepton/jet overlap removal selections of the analysis, we identify particle level quarks  with $p_T^j>30\;$GeV and $|\eta_j|<5$ as jets. In samples with more than one jet, we merge jets within $\Delta R = 0.4$. Moreover $\tau$ leptons are approximated as jets. 
Further selections are imposed on the kinematics of the event. In all SRs  the invariant mass of the SS lepton pairs must be $m_{\ell^\pm \ell^\pm}>200\;$GeV. Furthermore, in regions with more than two leptons events are rejected if any opposite-charge same-flavour lepton pair is within 10\;GeV of the $Z-$boson mass to remove diboson backgrounds.  Also, in sample with 2 or 3 leptons it is required $\Delta R(\ell^\pm \ell^\pm)<3.5$, 
$p_T^{\ell^\pm\ell^\pm}>100\;$GeV and $\sum|p_T^\ell|>300\;$GeV, where $p_T^{\ell^\pm\ell^\pm}$ is the vector sum of the SS lepton transverse momenta.
In the region with four leptons an additional selection is applied.
By defining $\bar M = \frac{m_{++}+m_{--}}{2}$ and $\Delta M = |m_{++}-m_{--}|$, the analysis requires $\Delta M$ values which are below $15-50\;$GeV for $\bar M$ = 200 GeV, $30-160\;$GeV for $\bar M$= 500 GeV, and $50-500\;$GeV for $\bar M$ = 1000 GeV, depending on the flavor of the leptons.\footnote{Since ATLAS does not provide any information on the flavor dependence of this cut, we take as cut value the average of the two values and linearly interpolate between different $\bar M$ values.} A summary of the selections is found in Tab.~3 of the ATLAS paper~\cite{ATLAS:2017xqs}.  From the measured number of events in each signal region and the estimation of the SM background we estimate the \textit{observed} number of signal events excluded at 95\% confidence level (CL) via the CLs procedure~\cite{Read:2000ru,Read:2002hq}. Projections for higher integrated luminosities are obtained by extrapolating the \textit{expected} 95\% CL exclusion. A summary of the exclusion yields are reported in Tab.~\ref{tab:atlas_excl} for the various SRs defined by the ATLAS search.

\begin{table}
\centering
\begin{tabular}{ c | c | c || c | c}
	Signal region	 &  $N_{\rm obs}$ 	& $N_{\rm bkg}$		& $N_{\rm obs}^{95\%CL}$ &  $N_{\rm exp}^{95\%CL}$	\\
	\hline
 $e^\pm e^\pm$ &  132	& 160$\pm 14$	&23.4	&39.0	\\
 $e^\pm \mu^\pm$ &  106	& 97.1$\pm 7.7$	&33.7	& 27.1	\\
  $\mu^\pm \mu^\pm$ &  26	& 22.6$\pm 2.0$	&	15.1&	12.4\\
\hline
  $e^\pm e^\pm e^\mp$ &  11	& 13.0$\pm 1.6$	&	8.1	& 9.7\\
  $e^\pm \mu^\pm \ell^\mp$ &  23	& 34.2$\pm 3.6$	&8.6	&16.0	\\
  $\mu^\pm \mu^\pm \mu^\mp$ &  13	& 13.2$\pm 1.3$	&	9.4& 10.2	\\
  $\ell^\pm \ell^\pm \ell^{\prime \mp}$ &  2	& 3.1$\pm 1.4$	&4.9	& 7.0	\\
  \hline
  $\ell^\pm\ell^\pm\ell^\mp\ell^\mp$ &  1	& 0.33$\pm 0.23$	&4.2	&4.2	\\
\end{tabular}
\caption{Number of observed events in the various signal region of the ATLAS search~\cite{ATLAS:2017xqs}, together with the SM background expectations. The observed and expected number of excluded signal events are computed via the CLs procedure~\cite{Read:2000ru,Read:2002hq}.}
\label{tab:atlas_excl}
\end{table}

\begin{figure}[t!]
\centering
\includegraphics[width=0.48\textwidth]{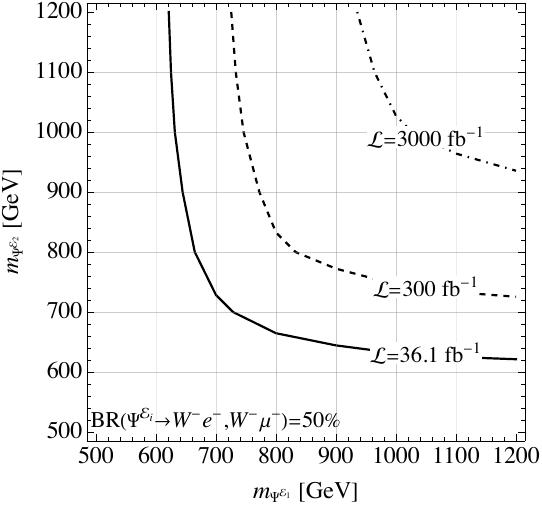}
\hfill
\includegraphics[width=0.48\textwidth]{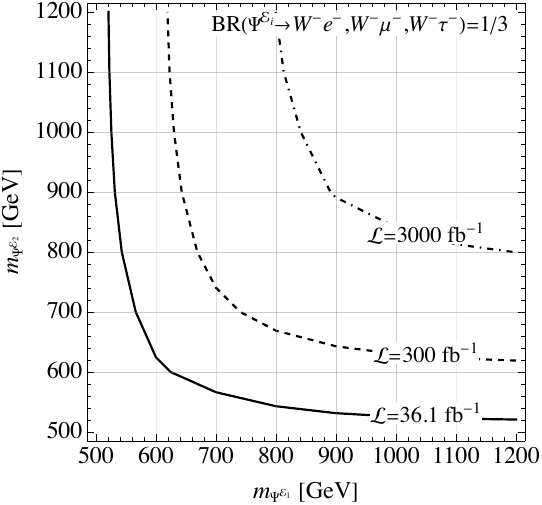}
\caption{
95\% CL exclusion limits obtained from the LHC ATLAS search~\cite{ATLAS:2017xqs}, solid line. The dashed and dot-dashed lines show the projected exclusion with $300\;$fb$^{-1}$ and \;3000\;fb$^{-1}$ of integrated luminosity respectively. In the left panel we assume exclusive mixing with the first two generation of SM leptons with equal weights, ${\rm BR}(\Psi^{\E_{1,2}} \to W^- e^-) = {\rm BR}(\Psi^{\E_{1,2}} \to W^- \mu^-) = 50\%$. In the right panel we assume them to democratically mix with all the three SM lepton families ${\rm BR}(\Psi^{\E_{1,2}} \to W^- e^-) = {\rm BR}(\Psi^{\E_{1,2}} \to W^- \mu^-) = {\rm BR}(\Psi^{\E_{1,2}} \to W^- \tau^-) = 1/3$.
}
\label{fig:ATLAS-exc}
\end{figure}

We then consider the pair-production of both the doubly charged leptons $p p \to \bar \Psi^{\E_{1,2}}\Psi^{\E_{1,2}}$ and study the limits obtained from the recast of the ATLAS search~\cite{ATLAS:2017xqs}, under various  assumptions for the flavor mixing pattern of the exotic leptons. For concreteness we study two scenarios. In the former both the exotic fermions only mix with the first two generations of SM leptons with equal weights, thus ${\rm BR}(\Psi^{\E_{1,2}} \to W^- e^-) = {\rm BR}(\Psi^{\E_{1,2}} \to W^- \mu^-) = 50\%$. In the latter we assume them to democratically mix with all the three SM lepton families ${\rm BR}(\Psi^{\E_{1,2}} \to W^- e^-) = {\rm BR}(\Psi^{\E_{1,2}} \to W^- \mu^-) = {\rm BR}(\Psi^{\E_{1,2}} \to W^- \tau^-) = 1/3$. The strongest exclusion limits come in both cases from SRs with three leptons and are shown in Fig.~\ref{fig:ATLAS-exc} for the actual luminosity of the ATLAS search, $\L = 36.1\;$fb$^{-1}$, as well as for two projected benchmark values, $\L = 300,\;3000\;$fb$^{-1}$. Current limits in the first scenario are around $600\;$GeV, if one of the two exotic lepton is decoupled from the spectrum, reaching $\sim 720\;$GeV if the two states are mass degenerate. At the end of the high-luminosity phase of the LHC, values up to $m_{\Psi^{\E_{1,2}}}\simeq 1\;$TeV could be tested. Allowing for a mixing with the $\tau$ lepton, relaxes these limits of ${\cal O}(100\;{\rm GeV})$ given that this final state is not directly targeted by the analysis~\cite{ATLAS:2017xqs}, and is anyway expected 
to be harder to be tested.

Turning our attention to the $Q=e$ exotic states, they promptly decay, for mixing angles $\gtrsim \mathcal{O}(10^{-7})$, as
\be
\Psi^{\N_i} \to W^- \nu \ , Z \ell^- \ ,
\ee
where the branching ratio of each channel is a function of the model parameters. The experimental signature depends on how the $W$ or $Z$ bosons decay and, in particular, signatures as in Tab.~\ref{tab:atlas_excl} are possible. The inclusion of the $\Psi^{\N_i}$ decay modes would then increase the constraints from direct searches, but it requires a less straightforward analysis than the one performed for the double charged states. We choose to be conservative and rely on the constraints we already obtained in Fig.~\ref{fig:ATLAS-exc} since the take-home message of this work, which is that purely chiral extensions of the SM lie beyond the edge of perturbativity, as we will discuss in the next section, would not substantially change.

\subsection{Perturbative unitarity}

In Sec.~\ref{sec:prompt} we have shown that current LHC limits push the exotic leptons to have a mass $\gtrsim 500-600\;$GeV, depending on the flavor structure of the model, in the case that they mix with the SM fermions. In case of no SM-BSM mixing these limits are pushed to 
even higher values, from the null results for searches of heavy-stable charged particles, see Sec.~\ref{sec:HSCP}.

If the exotic fermions are chiral states which acquire their mass only via their couplings to the Higgs, the Yukawa couplings $y_{1,2}$ and $y_{3,4}$ attain large values, possibly at the edge of the perturbative regime. In order to quantify this statement we apply the procedure discussed in~\cite{Allwicher:2021rtd}, see also \cite{DiLuzio:2016sur},
and compute the perturbative unitarity (PU) bounds obtained by considering all the $2\to 2$ scatterings present in the model. The strongest limit is obtained in the $J=0$ partial wave and, working for simplicity in the custodial limit
$y_1 = y_3$, $y_2 = y_4$, reads
\be
\label{eq:PUyuk}
3 |y_1|^2 + 3 |y_2|^2 + \sqrt{9 |y_2|^4 - 2 |y_1|^2 |y_2|^2 + 9 |y_2|^4} < 16\pi \ .
\ee
This bound can be then translated into a perturbative {\it upper limit} on the exotic fermion masses, which fixes $m_{\Psi^{\E_{1,2}}}\lesssim 400\;$GeV. Hence,  
current LHC limits  perturbatively exclude  the model, 
according to the criterium in \eq{eq:PUyuk}.


\section{Opening up the parameter space with vector-like masses} 
\label{sec:pheno_VL}


We have seen that SM extensions with exotic chiral fermions are pushed beyond the edge of their perturbativity regime by direct searches. In this section we thus consider the effect of the addition of vector-like masses, in order to avoid the constraints from direct searches while being compatible with the electroweak precision test and the recent measurement of $h\to Z\gamma$.

\subsection{Numerical analysis}

\begin{figure}[t!]
	\centering
	\includegraphics[width=0.46\textwidth]{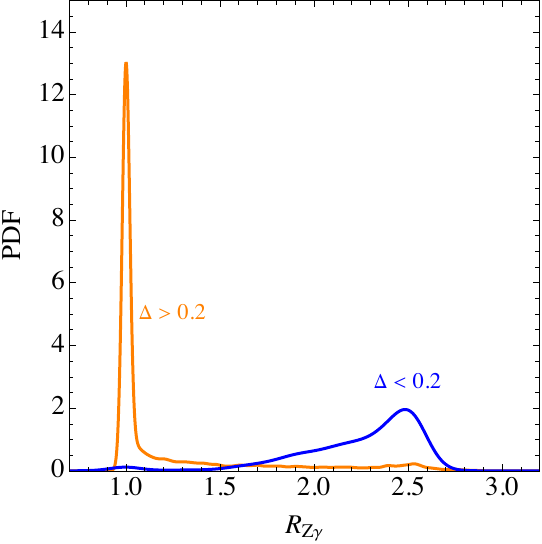}
	\hfill
	\includegraphics[width=0.49\textwidth]{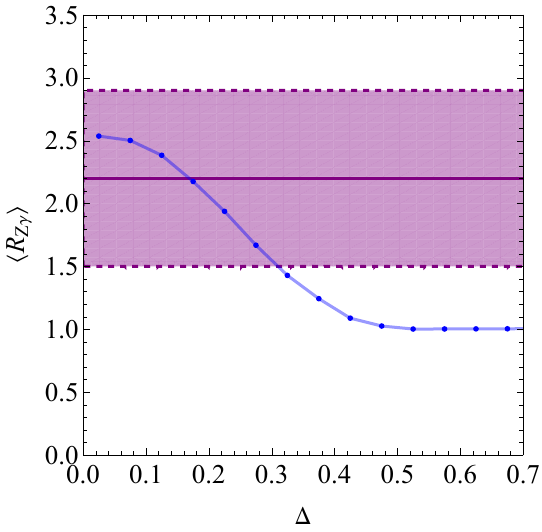}
	\caption{{\it Left panel:} Distribution of the $R_{Z\gamma}$ predictions for two regions of the mass-splitting parameters: $\Delta<0.2$ and $\Delta>0.2$.
 {\it Right panel:} Average value of the $R_{Z\gamma}$ predictions as  function of the mass-splitting parameter $\Delta$.}
	\label{plot:genVL}
\end{figure}

Since we were not able in general to cast explicit expressions for the $h\to\gamma\gamma,Z\gamma$ theoretical predictions and for the Peskin-Takeuchi parameters into a simple analytical form, see \eqs{eq:Agg}{eq:BZg} and \eqs{eq:Tgen}{eq:Sgen}, we rely on a numerical analysis to study the $R_{Z\gamma}$ prediction when the vector-like masses are included.
In particular, we randomly scan the parameters of the Lagrangian in Eq.~\eqref{eq:yukawa} and Eq.~\eqref{eq:VLmass} within the intervals\footnote{The parameters that allows mixing along exotic and SM leptons are highly constrained by precise measurements of the $Z$ coupling and hence are negligible in all the practical calculations.}
\be
|m_{\L,\,\N,\,\E}| \in \left[1.7,2\right] \ \text{TeV} 
\qquad \text{and} \qquad
|y_{1,\,2,\,3,\,4}|\in \left[1,\sqrt{\frac{8\pi}{5}}\right] \, ,
\ee
so that the PU bound on the Yukawa couplings is always satisfied. 
For each point extracted, the value of the hypercharge parameter $Y$ is set by minimizing the $\chi^2$ of the $S$ and $R_{\gamma\gamma}$ observables\footnote{The observable $T$ is independent of the hypercharge $Y$ and hence it is not considered in the $\chi^2$ analysis.}
\be
\chi^2=\left(\frac{R_{\gamma\gamma}-\mu_{\gamma\gamma}}{\sigma_{\gamma\gamma}}\right)^2+\left(\frac{S-\mu_{S}}{\sigma_{S}}\right)^2 \ ,
\ee
where $\mu_{\gamma\gamma}=1.00$, $\sigma_{\gamma\gamma}=0.12$, $\mu_{S}=0.05$, $\sigma_{S}=0.11$.
Finally we evaluate the theoretical prediction of the $h\to Z\gamma$ channel for each point and plot the results in \fig{plot:genVL}, where the parameter
\be
\Delta\equiv\left|\frac{m_{\Psi^{\E_{1}}}^2-m_{\Psi^{\E_{2}}}^2}{m_{\Psi^{\E_{1}}}^2+m_{\Psi^{\E_{2}}}^2}\right|+\left|\frac{m_{\Psi^{\N_{1}}}^2-m_{\Psi^{\N_{2}}}^2}{m_{\Psi^{\N_{1}}}^2+m_{\Psi^{\N_{2}}}^2}\right|
\ee
measures the degeneracy of the $\E$ and $\N$ exotic states. As we can see from the plots, the inclusion of vector-like mass terms would generally lead to $R_{Z\gamma}$ values around unity, {\it i.e.}~SM-like, unless the exotic leptons are degenerate, in which case the $h\to Z \gamma$ prediction lies within the $1\sigma$ region of the ATLAS and CMS combined measurement.

\subsection{Limit of degenerate masses}

\begin{figure}[t!]
	\centering
	\includegraphics[width=0.48\textwidth]{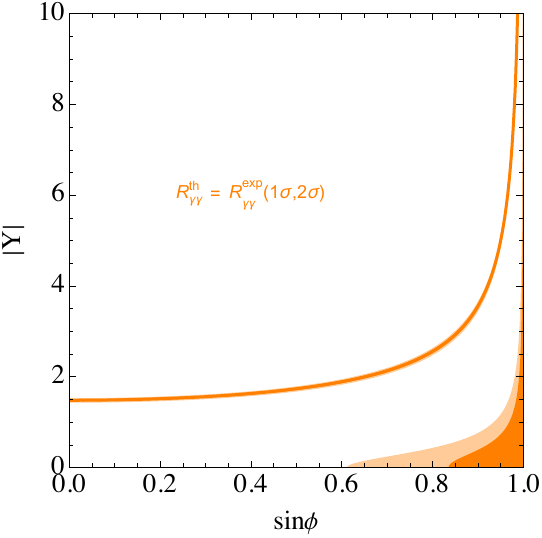}
	\hfill
	\includegraphics[width=0.48\textwidth]{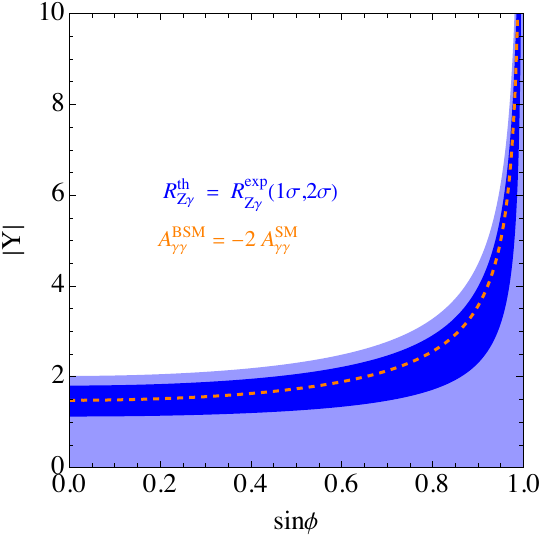}
	\caption{{\it Left panel}: Allowed region 
 of $R_{\gamma\gamma}$
 at $1\sigma$ and $2\sigma$ 
 in the $\sin\phi - |Y|$ plane.
 {\it Right panel}: Allowed region 
 of $R_{Z\gamma}$
 at $1\sigma$ and $2\sigma$  
 in the $\sin\phi - |Y|$ plane.
 The dashed yellow line indicates the non-decoupling solution $A_{\gamma\gamma}^{\text{BSM}}\simeq-2A_{\gamma\gamma}^{\text{SM}}$.}
	\label{plot:RVL}
\end{figure}

As shown above, the $h\to Z\gamma$ prediction naturally lies in the $1\sigma$ region of the ATLAS and CMS measurement in the limit of mostly degenerate masses, {\it i.e.}~$\Delta\to0$.
Motivated by this result, we now specify our study to the case of fully degenerate exotic masses, that is
\be
\label{eq:degeneratemass}
\M_\E^\dag\M_\E=\M_\N^\dag\M_\N=m_\Psi^2
\begin{pmatrix}
1 & 0 \\
0 & 1 \\
\end{pmatrix} \ ,
\ee
with $m_\Psi$ denoting the common mass value. \eq{eq:degeneratemass} enforces the mass matrices to be
\be
\M_\E=\M_\N=
\begin{pmatrix}
M & \pm i\frac{v}{\sqrt{2}}y_H \\
\pm i\frac{v}{\sqrt{2}}y_H & M \\
\end{pmatrix}
=m_\Psi
\begin{pmatrix}
\sin{\phi} & \pm i\cos{\phi} \\
\pm i\cos{\phi} & \sin{\phi} \\
\end{pmatrix} \ ,
\ee
with $y_H,M$ real and positive parameters, while the physical mass reads
\be
m_\Psi^2=M^2+\frac{1}{2}(y_Hv)^2
\ee
and 
\be
\cos{\phi}=\frac{1}{\sqrt{2}}\frac{y_Hv}{m_\Psi} \ ,
\ee
is the cosine of the angle measuring the relative contribution of the chiral mass on the total one.

The PU constraint \eq{eq:PUyuk} enforces on the degenerate case the bound $y_H^2 < 8 \pi/5$.
In the degenerate limit the Peskin-Takeuchi parameter $T$ is zero due to the custodial symmetry while
\be
S \simeq \frac{1}{3\pi}\cos^2\!\phi \ ,
\ee
which is always compatible with the experimental constraint for every value of the cosine.
Furthermore, the insertion of vector-like masses in the degenerate limit modifies \eq{A1} as 
\be
\label{A2}
\begin{split}
& A_{\gamma\gamma}^{\rm BSM} \simeq  \frac{4}{3}\left(1+  4 Y^2 \right)\cos^2\!{\phi} \ ,\\
& A_{Z\gamma}^{\rm BSM} \simeq  \frac{2}{3}\left[1 -   (1+8Y^2) {\rm tg}_w^2 \right]\cos^2\!{\phi} \  ,
\end{split}
\ee
so that $R_{\gamma\gamma,\, Z\gamma}$ depend on two parameters, the hypercharge $Y$ and mixing angle $\phi$.
In \fig{plot:RVL} we show the $1\sigma$, dark shaded, and $2\sigma$, light shaded, allowed regions of the $R_{\gamma\gamma,\, Z\gamma}$ observables 
in the $\sin\phi - |Y|$ plane.
Note that the $R_{\gamma\gamma}$ plot shows two different allowed regions, relative to the two SM-like solutions: the decoupling scenario $A_{\gamma\gamma}^{\text{BSM}}\simeq0$ in the bottom right corner of the figure and the non-decoupling one $A_{\gamma\gamma}^{\text{BSM}}\simeq-2A_{\gamma\gamma}^{\text{SM}}$.
Different considerations on these solutions then apply depending on if the 
exotic leptons are stable or unstable.

\begin{figure}[t!]
	\centering
	\includegraphics[width=0.48\textwidth]{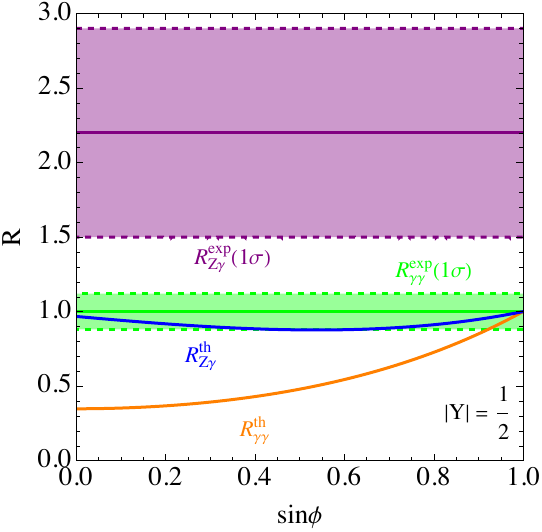}
	\hfill
	\includegraphics[width=0.48\textwidth]{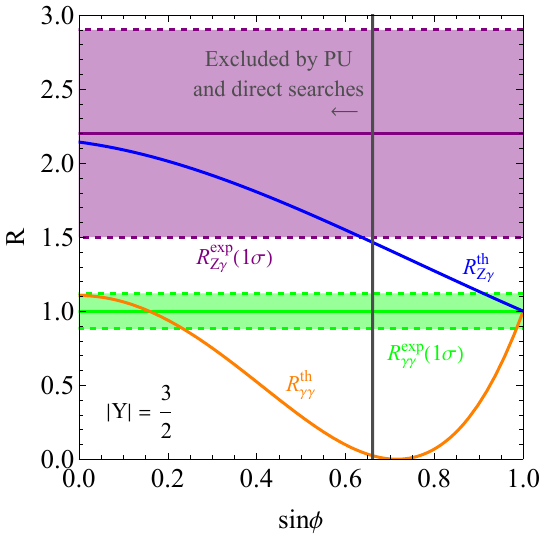}
	\caption{
 $R_{\gamma\gamma}$ and $R_{Z\gamma}$ signal strengths as a function of $\sin\phi$ for $|Y|=1/2$ (left panel) and 
 $\sin\phi$ for $|Y|=3/2$ (right panel).  The area between the shaded lines around the central value corresponds to the current $1\sigma$ experimental uncertainty. The region on the left of the gray line is excluded by the PU and direct search constraints discussed in Sec.~\ref{sec:HSCP}.
}
	\label{plot:stableVL}
\end{figure}

\subsubsection{Unstable exotic leptons}

Exotic leptons can mix with the SM fields, and hence decay into SM states, only for $|Y|=\frac{1}{2},\frac{3}{2}$. 
We plot in \fig{plot:stableVL} the $h\to\gamma\gamma,Z\gamma$ predictions 
for these two hypercharge values
as a function of $\sin\phi$.
For $|Y|=1/2$ we observe on left panel that the $h\to\gamma\gamma$ prediction is compatible with the experimental value only in the decoupling region where 
$R_{\gamma\gamma,Z\gamma}\approx1$.
A similar conclusion holds for $|Y|=3/2$, where we observe on the right panel that the PU bound on the Yukawa couplings and the direct search constraints discussed in Sec.~\ref{sec:HSCP} exclude the $A_{\gamma\gamma}^{\rm BSM}\approx-2A_{\gamma\gamma}^{\rm SM}$ solution, leading again to a scenario with decoupled exotic leptons. 
In either of the two cases above, it is then trivially possible 
in the decoupling regime
to be compatible with the SM prediction $R_{\gamma\gamma,Z\gamma}\approx1$. However, it is not possible to explain the 2$\sigma$ hint for $h\to Z\gamma$.

\subsubsection{Stable exotic leptons}

\begin{figure}[t!]
	\centering
	\includegraphics[width=0.49\textwidth]{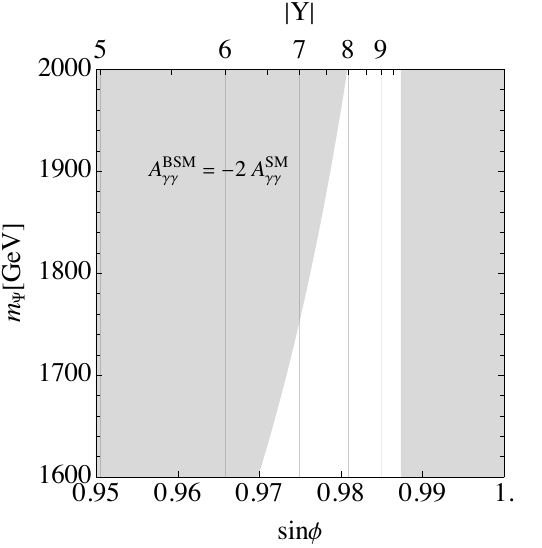}
	\hfill
	\includegraphics[width=0.475\textwidth]{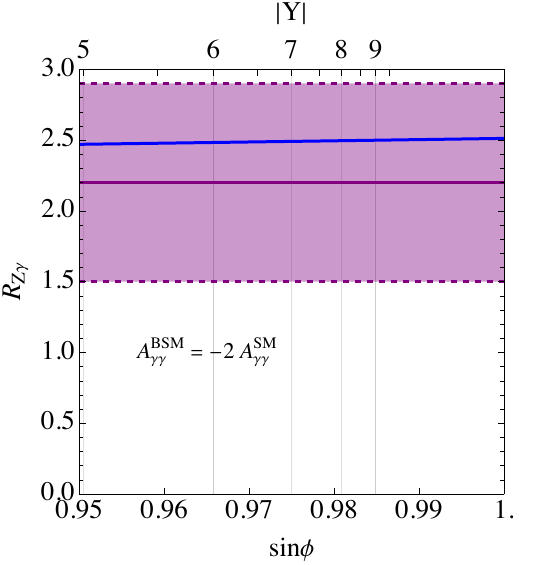}
	\caption{{\it Left panel}: Exclusion region due to PU bounds and direct searches constraints in the $\sin\phi- m_\Psi$ plane. {\it Right panel}: $R_{Z\gamma}$ as a function of $\sin\phi$. In both panel the corresponding value of $Y$ is also shown and the non-decoupling solution $A_{\gamma\gamma}^{\rm BSM}=-2A_{\gamma\gamma}^{\rm SM}$ is imposed.}
	\label{plot:PU}
\end{figure}

For $|Y|\neq1/2,3/2$ mixing terms with the SM fermions are forbidden and the exotic leptons are stable, hence direct searches constrain the exotic mass to be $\gtrsim1600$ GeV as already discussed in Sec.~\ref{sec:HSCP}.
Such strong constraint leads to a mostly-decoupled scenario where vector-like masses must dominate over the chiral contribution. Hence in order to obtain values of $R_{Z\gamma}$ different from unity, large values of the hypercharge $Y$ are required.
Along with the one on the Yukawa couplings, a PU bound on the hypercharge $Y$ can be computed with the procedure discussed in Ref.~\cite{Barducci:2023lqx} by considering the $\psi B \to\psi B$ scattering\footnote{This is the only $2\to2$ scattering channel with non-singular partial amplitude when the vector field is taken to be exactly massless in the high-energy limit.}, where $B$ is the hypercharge field and $\psi$ the exotic lepton field. The strongest limit is obtained in the $J=1/2$ partial wave and reads
\be
\text{Max}\Biggl\{Y^2, \ \left(Y+\frac{1}{2}\right)^2, \ \left(Y-\frac{1}{2}\right)^2\Biggl\} < \frac{4\pi}{g_Y^2}  \ ,
\ee
which yields $|Y| \lesssim 10$ for $g_Y=0.36$.
In the left panel of \fig{plot:PU} we show the exclusion region due to the PU bounds on the Yukawa couplings and the hypercharge $Y$ as a function of the degenerate mass $m_\Psi$ and the lower limit enforced by LHC direct searches, while imposing the non-decoupling solution $A_{\gamma\gamma}^{\rm BSM}=-2A_{\gamma\gamma}^{\rm SM}$. In the region allowed by PU and direct search constraints, the $R_{Z\gamma}$ prediction is around $\sim2.5$ as shown in the right panel of \fig{plot:PU}.

In the near future the HL-LHC is expected to measure the $h\to Z\gamma$ decay within 10$\%$ precision \cite{Cepeda:2019klc}. In particular ATLAS plans to reduce the uncertainty from the present value $\pm0.9$ \cite{CMS:2023mku} down to $\pm0.23$ \cite{Cepeda:2019klc}, hence to improve it by a factor of $\sim25\%$.
By considering that the same improvement will be obtained for the uncertainty of a combined 
ATLAS and CMS
analysis and assuming that the present 
central value will not change with future measurements, one will obtain 
$R_{Z\gamma}^{\text{HL-LHC}}=2.2\pm0.2$, that is a measured value which is more than $5\sigma$ away from the SM value.
In such a, highly speculative, case one would 
need some non-decoupling new physics, 
like the one 
discussed in the present work, 
in order to keep $R_{\gamma\gamma}$ SM-like and 
explain at the same time a non-standard value of $R_{Z\gamma}$. 







\section{Conclusions}
\label{sec:concl}

In this work we presented a detailed exploration of the phenomenology associated with chiral extensions to the SM 
coupled to a single Higgs doublet, particularly focussing on exotic leptons. Our study revisits and expands upon the original analysis provided in Ref.~\cite{Bizot:2015zaa}, incorporating the latest data from LHC direct searches. These insights have been crucial in delineating the parameter space for these chiral extensions, especially in light of the stringent constraints imposed by recent experimental results.

Our findings highlight that purely chiral scenarios are perturbatively ruled out when considering the combined constraints from Higgs coupling measurements and direct LHC searches. On the other hand, by incorporating vector-like mass components into the exotic leptons, we observe an intriguing expansion of the viable parameter space. This extended framework not only aligns with the current 2$\sigma$ deviation observed in the 
$h \to Z\gamma$ channel but also provides a fertile ground for future experimental investigations, particularly with the upcoming high-luminosity phase of the LHC.

As a by-product of our analysis we also conclude that the possibility pointed out in Ref.~\cite{DiLuzio:2022ziu} to avoid the strong constraints stemming from 
the coupling of light vector bosons to anomalous SM currents 
\cite{Dror:2017ehi,Dror:2017nsg}, 
is not anymore viable, at least perturbatively, 
since it relies on the hypothesis 
of purely chiral anomalon fields. 




\section*{Acknowledgments}
 
We thank Michele Frigerio for useful discussions. 
The work of LDL, MN and CT 
is supported by the 
by the Italian Ministry of University and Research (MUR) 
via the PRIN 2022 project n.~2022K4B58X -- AxionOrigins. 
The work of LDL is also supported by the European Union -- NextGenerationEU and by the University of Padua under the 2021 STARS Grants@Unipd programme (Acronym and title of the project: CPV-Axion -- Discovering the CP-violating axion), 
the European Union's Horizon 2020 research and innovation programme under the 
Marie Sk\l{}odowska-Curie grant agreement n.~101086085 -- ASYMMETRY
and the INFN Iniziative Specifica APINE.

\appendix

\section{Other chiral extensions of the Standard Model}
\label{app1}

In this Appendix we consider more general chiral extensions of the SM, other than the one of \eq{eq:fermions_QN}, and show that they do not allow to reproduce a value of $R_{Z\gamma}$ compatible with the current experimental limits, thus supporting the choice of the field content studied in the main text.

\subsection{Classification with up to four chiral multiplets}

Let us consider fermionic extensions of the SM given by an 
anomaly-free set of up to four chiral multiplets, 
as discussed in \cite{Bizot:2015zaa}.
We require that no vector-like mass terms are allowed, no exotic state is left massless after EWSB and that each field is a singlet under QCD.
These requirements leave us with only one possible set, composed by four chiral BSM fermions transforming under the SM as \cite{Bizot:2015zaa}
\be\label{eq:fermions_QN3}
\chi_{2R}\sim ({\textbf{1}},{\textbf{2}}I{\textbf{+2}})_0 \ , \quad 
\psi_{1R} \sim ({\textbf{1}},{\textbf{2}}I{\textbf{+1}})_{-\frac{1}{2}} \ , \quad 
\psi_{2R} \sim ({\textbf{1}},{\textbf{2}}I{\textbf{+1}})_{+\frac{1}{2}} \ , \quad
\chi_{1R} \sim ({\textbf{1}},{\textbf{2}}I)_{0} \ ,
\ee
with the isospin quantum number $I$ being necessarily a semi-integer to avoid the global $SU(2)_L$ anomaly \cite{Witten:1982fp}.
All the fields can take mass from the SM Higgs through the Yukawa Lagrangian
\be\label{lag:an3}
\begin{split}
-{\cal L}_{\text{Yukawa}}= &y_{21}\bar{(\psi_{2R})^c}\tilde{H}\chi_{1R} + y_{22}\bar{(\psi_{2R})^c}\tilde{H}\chi_{2R} \\
+& y_{11}\bar{(\chi_{1R})^c}H\psi_{1R} + y_{12}\bar{(\chi_{2R})^c}H\psi_{1R} \,+\,h.c. \, ,
\end{split}
\ee
where isospin contractions are left implicit.
The Lagrangian of \eq{lag:an3} admits the global symmetry
\begin{equation}
U(1)_R: \qquad \chi_{iR}\to e^{i\theta}\chi_{iR} \, , \,\, \psi_{iR}\to e^{-i\theta}\psi_{iR} \qquad i=1,2 \, ,  
\end{equation}
that can prevent any additional Yukawa or mass term that involves the exotic fields.
Assuming the exotic fermions to be much heavier than the Higgs and $Z$ bosons after EWSB, one gets the following BSM contribution to $h\to\gamma\gamma,Z\gamma$ decay
\begin{align}
A_{\gamma\gamma}^{\rm BSM}\simeq& \frac{4}{9}(2I+1)(3+4I+4I^2)\, , \\
A_{Z\gamma}^{\rm BSM}\simeq& \frac{1+2I}{9}[3+8I+8I^2-{\rm tg}_w^2(9+8I+8I^2)] \, .
\end{align}
In Tab.~\ref{tab:app1} we report the $R_{\gamma\gamma,Z\gamma}$ predictions for the lowest values of the isospin $I$. As one can note, we are never able to reproduce the experimental results for 
any value of $I$.

\begin{table}[h!]
\begin{center}
\begin{tabular}{c||c|c|c}
 & $I=\frac{1}{2}$ & $I=\frac{3}{2}$ & $I=\frac{5}{2}$ \\
\hline\hline
$R_{\gamma\gamma}$ & $\sim0.03$ & $\sim15$ & $\sim213$ \\
\hline
$R_{Z\gamma}$ & $\sim0.7$ & $\sim0.3$ & $\sim19$ \\
\end{tabular}
\end{center}
\caption{$R_{\gamma\gamma,Z\gamma}$ predictions for the lowest values of the isospin $I$.}
\label{tab:app1}
\end{table}

\subsection{Generalization to higher-dimensional $SU(2)_L$ representations}

\begin{figure}[t!]
	\centering
	\includegraphics[width=0.48\textwidth]{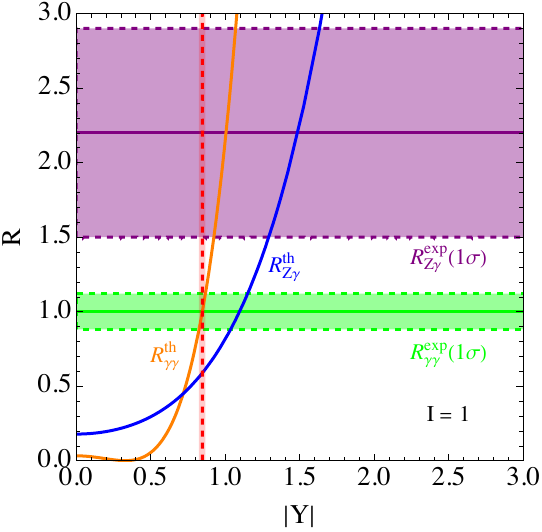}
	\caption{
    $R_{\gamma\gamma}$ and $R_{Z\gamma}$ signal strengths as a function of $|Y|$ for $I=1$. The area between the shaded lines around the central value corresponds to the current $1\sigma$ experimental uncertainty. The red, vertical band denotes the values of $Y$ able to satisfy the di-photon constraints at 1$\sigma$.}
	\label{plotI}
\end{figure}

We next explore a generalization of the field content described in \eq{eq:fermions_QN} with arbitrary $SU(2)_L$ representations.
In particular, we consider BSM fermions vector-like under the SM, with respect to which they transform as
\be\label{eq:fermions_QN2}
\L\sim ({\textbf{1}},{\textbf{2}}I+{\textbf{1}})_Y \ , \quad 
\E \sim ({\textbf{1}},{\textbf{2}}I)_{Y-\frac{1}{2}} \ , \quad 
\N \sim ({\textbf{1}},{\textbf{2}}I)_{Y+\frac{1}{2}} \ , \quad
\F \sim ({\textbf{1}},{\textbf{2}}I-{\textbf{1}})_{Y} \ ,
\ee
where $I$ is the isospin of the largest $SU(2)_L$ representation, whose value is restricted to integer or semi-integer values. Note that for $I=1/2$ the $\F$ field is absent and we are left with the same field content as \eq{eq:fermions_QN}.
All the fields can take mass from the SM Higgs through the Yukawa Lagrangian
\be\label{lag:an2}
\begin{split}
-{\cal L}_{\text{Yukawa}}=&y_{R1}\bar{\L}_L\tilde{H}\N_R + y_{L1}\bar{\L}_R\tilde{H}\N_L + y_{R2}\bar{\F}_L\tilde{H}\N_R + y_{L2}\bar{\F}_R\tilde{H}\N_L \\
+& y_{R3}\bar{\L}_LH\E_R + y_{L3}\bar{\L}_RH\E_L + y_{R4}\bar{\F}_LH\E_R + y_{L4}\bar{\F}_RH\E_L \,+\,h.c. \, ,
\end{split}
\ee
where isospin contractions are left implicit.
The Lagrangian in \eq{lag:an2} admits two global symmetries, which are
\begin{align}
U(1)_V:& \, \L\to e^{i\alpha}\L \, , \,\, \E\to e^{i\alpha}\E \, , \,\, \N\to e^{i\alpha}\N \, , \,\, \F\to e^{i\alpha}\F \, , \\
U(1)_A:& \, \L\to e^{i\beta\gamma_5}\L \, , \,\, \E\to e^{-i\beta\gamma_5}\E \, , \,\, \N\to e^{-i\beta\gamma_5}\N \, , \,\, \F\to e^{i\beta\gamma_5}\F \, ,
\end{align}
that can prevent any additional Yukawa or mass term that involves the exotic fields. Let us consider the case in 
which no other mass term or mass mixing with the SM fields is present. 
Assuming the exotic fermions to be much heavier than the Higgs and $Z$ bosons after EWSB, one gets the following BSM contribution to $h\to\gamma\gamma,Z\gamma$ decays
\begin{align}
A_{\gamma\gamma}^{\rm BSM}\simeq& \frac{16}{9}I(1+2I^2+6Y^2)\, , \\
A_{Z\gamma}^{\rm BSM}\simeq& \frac{4}{9}I(1+{\rm tg}_w^2)(1+8I^2)-\frac{16}{9}I(1+2I^2+6Y^2){\rm tg}_w^2\, .
\end{align}
For $I\geq3/2$ the di-photon theoretical prediction is above the experimental result for all the values of $Y$ so we are left with only two relevant cases, which are $I=1/2$, already investigated in the main text, and $I=1$. We plot the 1$\sigma$ experimental regions and theoretical predictions of $R_{\gamma\gamma,Z\gamma}$ for $I=1$ in \fig{plotI}.
Note that for $I=1$, one gets
$Y=0.848_{-0.023}^{+0.021}$ 
from the requirement of matching the $R_{\gamma\gamma}$ result at $1\sigma$, thus leading to the prediction
$R_{Z\gamma}=0.588_{-0.028}^{+0.026}$,
which is smaller than the SM value and in contrast with the recent experimental osbervation.

\section{General formulae for 
loop-induced Higgs couplings and oblique corrections}
\label{app2}

In this Appendix
we present the calculation of the effective loop-induced Higgs couplings to gauge bosons that 
arise after integrating 
out heavy fermionic degrees of freedom and the contribution of the latter to the Peskin-Takeuchi parameters $S$ and $T$.
In particular, we focus on the 3-point vertices $hVV^\prime$ when gauge invariance prevents renormalizable couplings to the Higgs field, such that the tree level amplitudes are zero and one-loop amplitudes are free from UV divergences.

\subsection{General setup}

We assume a general setup with a set of gauge bosons $G_{\mu}^{A}$ 
related to 
the generators 
$Q^{A}$ 
of the gauge symmetry group $\mathcal{G}$, that can be in general semi-simple. 
The model contains a fermionic sector, whose fields are labeled as $\psi_{i}$, that acquire a mass term $\mathcal{M}_{ij}$ after a spontaneous symmetry breaking mechanism. The $(\frac{1}{2},0)$ and $(0,\frac{1}{2})$ Lorentz components of the $\psi$ field are separately reducible representations of 
$\mathcal{G}$
and the generators act on them as
\beq
Q^{A}\psi_{i}=\sum_{j}(Q_{L}^{A})_{ij}\psi_{jL}+\sum_{j}(Q_{R}^{A})_{ij}\psi_{jR} \, ,
\eeq
where $(Q_{L,R}^{A})_{ij}$ are the matrix rapresentation of the gauge multiplets $\psi_{L,R}\equiv P_{L,R}\psi$. 
We restrict ourselves to models with a $\U(1)_{\psi}$ symmetry corresponding to the fermionic number of the 
$\psi$ fields, $\psi_{i}\to e^{i\phi}\psi_{i}$.
The real scalar Higgs fields, responsible for the spontaneous symmetry breaking mechanism, are labeled as $H_{a}=(H_{a})^{*}$ 
and belong to a reducible representation of the gauge group $\mathcal{G}$. 
By performing an infinitesimal transformation of angle $\alpha_{A}$ along the $Q^{A}$ generator, the $H_{a}$ fields transform like
\beq
\delta H_{a}=\sum_{b}g_{A}\alpha_{A}(iQ_{H}^{A})_{ab}H_{b} \, ,
\eeq
where $(iQ_{H}^{A})_{ab}$ is a real and antisymmetric matrix.
Hence,  
\beq
\mathcal{L} \supset\sum_{i}\bar{\psi}_{i}i\slashed{\partial}\psi_{i}-\sum_{a,i,j}H_{a}(\bar{\psi}_{iL}\mathcal{Y}_{ij}^{a}\psi_{jR}+\text{h.c.})-\sum_{A}g_{A}G_{\mu}^{A}J^{\mu A} \, ,
\eeq
with
\beq
J^{\mu A}=\sum_{i,j}\Bigl[\bar{\psi}_{iL}\gamma^{\mu}(Q_{L}^{A})_{ij}\psi_{jL}+\bar{\psi}_{iR}\gamma^{\mu}(Q_{R}^{A})_{ij}\psi_{jR}\Bigr] \, .
\eeq
The Yukawa couplings must preserve gauge invariance and hence they satisfy
\beq
\sum_{k}\mathcal{Y}_{ik}^{a}(Q_{R}^{A})_{kj}-\sum_{k}(Q_{L}^{A})_{ik}\mathcal{Y}_{kj}^{a}+\sum_{b}\mathcal{Y}_{ij}^{b}(Q_{H}^{A})_{ba}=0 \, .
\eeq
The Higgs fields acquire the VEVs $\langle H_{a}\rangle=v_{a}$ which break the gauge group, leaving 
an unbroken subgroup $\mathcal{G}_{0}$. Then, the mass matrix of the $\psi$ fields is given by
\beq
\mathcal{M}_{ij}=\sum_{a}\mathcal{Y}_{ij}^{a}v_{a} \, ,
\eeq
leading to
\beq
\begin{split}
\mathcal{L}\supset&\sum_{i}\bar{\psi}_{i}i\slashed{\partial}\psi_{i}-\sum_{i,j}(\bar{\psi}_{iL}\mathcal{M}_{ij}\psi_{jR}+\text{h.c.})\\
&-\sum_{a,i,j}\tilde{H}_{a}(\bar{\psi}_{iL}\mathcal{Y}_{ij}^{a}\psi_{jR}+\text{h.c.})-\sum_{A}g_{A}G_{\mu}^{A}J^{\mu A} \, ,
\end{split}
\eeq
where $\tilde{H}_{a}=H_{a}\!-\!v_{a}$ 
are the Higgs fluctuations around the vacuum.

In order to go in the mass basis, 
the mass matrix $\mathcal{M}$ is diagonalized via the bi-unitary transformations 
$\psi_{R} \to U_{R} \, \psi_{R}$ 
and
$\psi_{L} \to U_{L} \, \psi_{L}$, 
which by construction satisfy $U_{L}^{\dag}\mathcal{M}U_{R}=
\text{diag}(m_{1},m_{2},...)$. This yields
\beq
\mathcal{L}\supset\sum_{i}\bar{\psi}_{i}(i\slashed{\partial}-m_{i})\psi_{i}-\sum_{A}g_{A}G_{\mu}^{A}J_{U}^{\mu A}-\sum_{a,i,j}\tilde{H}_{a}\bar{\psi}_{i}(\hat{\mathcal{Y}}_{R}^{a}P_{R}+\hat{\mathcal{Y}}_{L}^{a}P_{L})_{ij}\psi_{j} \, ,
\eeq
where $\hat{\mathcal{Y}}_{R}^{a}=U_{L}^{\dag}\mathcal{Y}^{a}U_{R}=(\hat{\mathcal{Y}}_{L}^{a})^{\dag}$, 
while the gauge currents in the mass basis are equal to
\beq
J_{U}^{\mu A}=\sum_{i,j}\Bigl[\bar{\psi}_{iL}\gamma^{\mu}(U_{L}^{\dag}Q_{L}^{A}U_{L})_{ij}\psi_{jL}+\bar{\psi}_{iR}\gamma^{\mu}(U_{R}^{\dag}Q_{R}^{A}U_{R})_{ij}\psi_{jR}\Bigr] \, .
\eeq

\subsection{Loop-induced Higgs couplings}

After integrating out the heavy fermion fields, we get 
effective operators of the type\footnote{We fix $\epsilon^{0123}=1$.}
\beq
\begin{split}
\label{eq:Ltoyeff}
\mathcal{L}\supset
&\sum_{a,B,C}C_{aBC}\tilde{H}_{a}(\partial_{\alpha}G_{\mu}^{B}-\partial_{\mu}G_{\alpha}^{B})(\partial_{\alpha}G_{\mu}^{C}-\partial_{\mu}G_{\alpha}^{C})\\
-&\sum_{a,B,C}D_{aBC}\epsilon^{\mu\nu\alpha\beta}\tilde{H}_{a}(\partial_{\alpha}G_{\mu}^{B}-\partial_{\mu}G_{\alpha}^{B})(\partial_{\beta}G_{\nu}^{C}-\partial_{\nu}G_{\beta}^{C})\, ,
\end{split}
\eeq
in terms of the effective field theory coefficients 
$C_{aBC}=C_{aCB}$ and $D_{a BC}=D_{a CB}$ 
that we want to compute.

\subsubsection{1-loop matching}

\begin{figure}[ht]
\centering
\includegraphics[width=13cm]{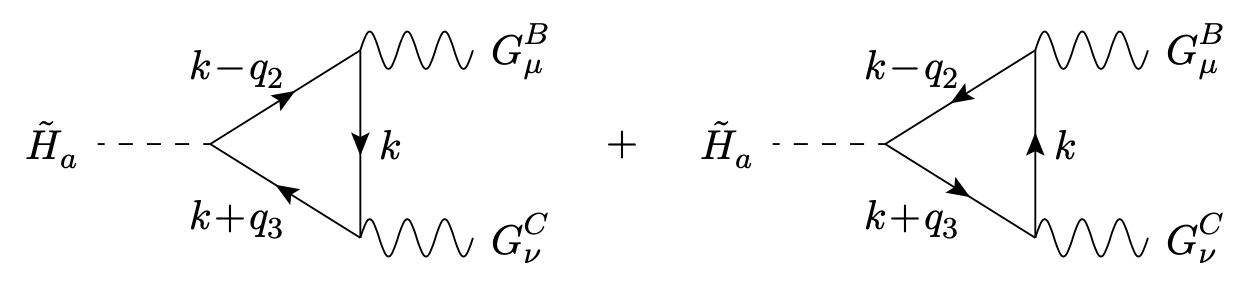}
\caption{Feynman diagrams relative to the 3-point functions in \eq{eq:MaBC}. }
\label{fig:dia3gauge}       
\end{figure}

Loop-induced Higgs coupling occurs in the 3-point functions $\Gamma_{ABC}^{\alpha\mu\nu}(x,y,z)$ and $\Gamma_{a BC}^{\mu\nu}(x,y,z)$ at 1-loop through fermionic triangle diagrams, see \fig{fig:dia3gauge}. The amplitudes in momentum space are defined via
\beq
\label{eq:MaBC}
\int\!\text{d}^{4}\!x\,\text{d}^{4}\!y\,\text{d}^{4}\!z\,e^{i(xq_{1}+yq_{2}+zq_{3})}\,\Gamma_{a BC}^{\mu\nu}(x,y,z)|_{\text{1-loop}}=(2\pi)^{4}\delta^{(4)}\!(q_{1}+q_{2}+q_{3})\,\mu^{\frac{4-d}{2}}\,iM_{a BC}^{\mu\nu}(q_{1},q_{2},q_{3}) \, .
\eeq
which yield
\beq
\begin{split}
M_{a BC}^{\mu\nu}=\sum_{\substack{i,j,k\\ \chi_{1},\chi_{2},\chi_{3}}}\!&g_{B}g_{C}(\hat{\mathcal{Y}}_{\chi_{1}}^{a})_{jk}(U_{\chi_{2}}^{\dag}Q_{\chi_{2}}^{B}U_{\chi_{2}})_{ki}(U_{\chi_{3}}^{\dag}Q_{\chi_{3}}^{C}U_{\chi_{3}})_{ij}\\
&\times i\mu^{4-d}\!\int\!\frac{\text{d}^{d}\!k}{(2\pi)^{d}}\,\frac{\text{Tr}_{\text{D}}[\gamma^{\mu}\!P_{\chi_{2}}\!(\slashed{k}+m_{i})\gamma^{\nu}\!P_{\chi_{3}}\!(\slashed{k}+\slashed{q}_{3}+m_{j})P_{\chi_{1}}\!(\slashed{k}-\slashed{q}_{2}+m_{k})]}{[k^{2}-m_{i}^{2}]\,[(k+q_{3})^{2}-m_{j}^{2}]\,[(k-q_{2})^{2}-m_{k}^{2}]}\\
+\sum_{\substack{i,j,k\\ \chi_{1},\chi_{2},\chi_{3}}}\!&g_{B}g_{C}(\hat{\mathcal{Y}}_{\chi_{1}}^{a})_{kj}(U_{\chi_{3}}^{\dag}Q_{\chi_{3}}^{C}U_{\chi_{3}})_{ji}(U_{\chi_{2}}^{\dag}Q_{\chi_{2}}^{B}U_{\chi_{2}})_{ik}\\
&\times i\mu^{4-d}\!\int\!\frac{\text{d}^{d}\!k}{(2\pi)^{d}}\,\frac{\text{Tr}_{\text{D}}[\gamma^{\nu}\!P_{\chi_{3}}\!(\slashed{k}+m_{i})\gamma^{\mu}\!P_{\chi_{2}}\!(\slashed{k}+\slashed{q}_{2}+m_{k})P_{\chi_{1}}\!(\slashed{k}-\slashed{q}_{3}+m_{j})]}{[k^{2}-m_{i}^{2}]\,[(k+q_{2})^{2}-m_{k}^{2}]\,[(k-q_{3})^{2}-m_{j}^{2}]}\,,
\end{split}
\eeq
where we have regularized the theory with dimensional regularization. One finds that the $q_3^\mu q_2^\nu$ and $\epsilon^{\mu\nu\alpha\beta}$ terms we are interested in are finite, {\it i.e.}~they do not contain 
$1/(d\!-\!4)$ poles, and are independent from the renormalization scale $\mu$\footnote{Instead, the term proportional to $g^{\mu\nu}$ could be affected by a divergence if the vertex is allowed at tree level but we assume the presence of a unbroken gauge symmetry that forbid the tree-level vertex and hence resolve the divergence. So we only focus on the finite terms of the 1-loop amplitude.}. Hence, we are free to send $d\rightarrow4$.
In order to obtain the effective field theory coefficients in \eq{eq:Ltoyeff}, we have to match the expressions that we have calculated above 
to the effective field theory matrix elements in the limit of heavy fermion masses, i.e.
\beq
\lim_{\substack{m_{i,j,k}^{2}\gg \\ q_{2}^{2},q_{3}^{2},q_{2}\!\cdot\!q_{3}}}M_{a BC}^{\mu\nu}=4C_{aBC}[q_{3}^\mu q_{2}^\nu-(q_3\cdot q_2)g^{\mu\nu}]+8D_{a BC}\epsilon^{\mu\nu\alpha\beta}q_{2\alpha}q_{3\beta}\,.
\eeq
Finally we get
\beq
\label{eq:Cgeneral}
\begin{split}
C_{aBC}=\frac{g_Bg_C}{32\pi^2}\int_{0}^{+\infty}\!\!\!\!\!\!\text{d}s\!\int_{0}^{1}\!\text{d}x\!\int_{0}^{1}\!&\text{d}y\!\int_{0}^{1}\!\text{d}z\,2\,\delta(1\!-\!x\!-\!y\!-\!z)\times\\
\times\,\text{Re}\Biggl\{ x\,&\text{Tr}\!\left[\Y^ae^{-sz\mathcal{M}^{\dag}\mathcal{M}}Q_{R}^{B}e^{-sx\mathcal{M}^{\dag}\mathcal{M}}\M^\dag Q_{L}^{C}e^{-sy\mathcal{M}\mathcal{M}^\dag}\right]\\
+x\,&\text{Tr}\!\left[\Y^ae^{-sz\mathcal{M}^{\dag}\mathcal{M}}Q_{R}^{C}e^{-sx\mathcal{M}^{\dag}\mathcal{M}}\M^\dag Q_{L}^{B}e^{-sy\mathcal{M}\mathcal{M}^\dag}\right]\\
+y(1-2z)\,&\text{Tr}\!\left[\Y^ae^{-sz\mathcal{M}^{\dag}\mathcal{M}}Q_{R}^{B}e^{-sx\mathcal{M}^{\dag}\mathcal{M}}Q_{R}^{C}\M^\dag e^{-sy\mathcal{M}\mathcal{M}^\dag}\right]\\
+y(1-2z)\,&\text{Tr}\!\left[\Y^ae^{-sz\mathcal{M}^{\dag}\mathcal{M}}Q_{R}^{C}e^{-sx\mathcal{M}^{\dag}\mathcal{M}}Q_{R}^{B}\M^\dag e^{-sy\mathcal{M}\mathcal{M}^\dag}\right]\\
+z(1-2y)\,&\text{Tr}\!\left[\Y^ae^{-sz\mathcal{M}^{\dag}\mathcal{M}}\M^\dag Q_{L}^{B}e^{-sx\mathcal{M}\mathcal{M}^{\dag}}Q_{L}^{C} e^{-sy\mathcal{M}\mathcal{M}^\dag}\right]\\
+z(1-2y)\,&\text{Tr}\!\left[\Y^ae^{-sz\mathcal{M}^{\dag}\mathcal{M}}\M^\dag Q_{L}^{C}e^{-sx\mathcal{M}\mathcal{M}^{\dag}}Q_{L}^{B}e^{-sy\mathcal{M}\mathcal{M}^\dag}\right]\Biggr\} \, ,
\end{split}
\eeq
and
\beq
\label{eq:Dgeneral}
\begin{split}
D_{a BC}=\frac{g_Bg_C}{64\pi^2}\int_{0}^{+\infty}\!\!\!\!\!\!\text{d}s\!\int_{0}^{1}\!\text{d}x\!\int_{0}^{1}\!&\text{d}y\!\int_{0}^{1}\!\text{d}z\,2\,\delta(1\!-\!x\!-\!y\!-\!z)\times\\
\times\,\text{Im}\Biggl\{x\,&\text{Tr}\!\left[e^{-sz\mathcal{M}^{\dag}\mathcal{M}}Q_{R}^{B}\mathcal{M}^{\dag}e^{-sx\mathcal{M}\mathcal{M}^{\dag}}Q_{L}^{C}e^{-sy\mathcal{M}\mathcal{M}^{\dag}}\mathcal{Y}^{a}\right]\\
+x\,&\text{Tr}\!\left[e^{-sz\mathcal{M}^{\dag}\mathcal{M}}Q_{R}^{C}\mathcal{M}^{\dag}e^{-sx\mathcal{M}\mathcal{M}^{\dag}}Q_{L}^{B}e^{-sy\mathcal{M}\mathcal{M}^{\dag}}\mathcal{Y}^{a}\right]\\
+y\,&\text{Tr}\!\left[e^{-sz\mathcal{M}^{\dag}\mathcal{M}}Q_{R}^{B}e^{-sx\mathcal{M}^{\dag}\mathcal{M}}Q_{R}^{C}\mathcal{M}^{\dag}e^{-sy\mathcal{M}\mathcal{M}^{\dag}}\mathcal{Y}^{a}\right]\\
+y\,&\text{Tr}\!\left[e^{-sz\mathcal{M}^{\dag}\mathcal{M}}Q_{R}^{C}e^{-sx\mathcal{M}^{\dag}\mathcal{M}}Q_{R}^{B}\mathcal{M}^{\dag}e^{-sy\mathcal{M}\mathcal{M}^{\dag}}\mathcal{Y}^{a}\right]\\
+z\,&\text{Tr}\!\left[e^{-sz\mathcal{M}^{\dag}\mathcal{M}}\mathcal{M}^{\dag}Q_{L}^{C}e^{-sx\mathcal{M}\mathcal{M}^{\dag}}Q_{L}^{B}e^{-sy\mathcal{M}\mathcal{M}^{\dag}}\mathcal{Y}^{a}\right]\\
+z\,&\text{Tr}\!\left[e^{-sz\mathcal{M}^{\dag}\mathcal{M}}\mathcal{M}^{\dag}Q_{L}^{B}e^{-sx\mathcal{M}\mathcal{M}^{\dag}}Q_{L}^{C}e^{-sy\mathcal{M}\mathcal{M}^{\dag}}\mathcal{Y}^{a}\right]\Biggr\}\,.
\end{split}
\eeq

\subsubsection{Application to $h\to\gamma\gamma,Z\gamma$}

Our results can be applied to the case of $h\to\gamma\gamma,Z\gamma$ if $B=\gamma$ and $C=Z$, for which
\be
g_\gamma=e \, , \qquad g_Z=\frac{g}{c_w} \, , \qquad Q^\gamma=Q \qquad \text{and} \qquad Q^Z=T^3-s_w^2Q \, ,
\ee
where $T^3$ is the diagonal isospin and $Q$ is the electric charge.
By defining
\be
R_{\gamma \gamma, Z \gamma} = \frac{|A_{\gamma\gamma,Z\gamma}^{\rm SM}+A_{\gamma\gamma,Z\gamma}^{\rm BSM}|^2+|\tilde{A}_{\gamma\gamma,Z\gamma}^{\rm BSM}|^2}{|A_{\gamma\gamma,Z\gamma}^{\rm SM}|^2} \, ,
\ee
with $A_{\gamma\gamma}^{\rm SM} \simeq -6.5  $, $A_{Z\gamma}^{\rm SM} \simeq -6.64$, one gets
\begin{gather}
\label{eq:Agg}
A_{\gamma\gamma}^{\rm BSM} \simeq \frac{2}{3}v\,\text{Re}\Biggl\{ \text{Tr}\left[ Q_R^2\M^{-1}\Y^h\right]
+ \text{Tr}\left[ Q_L^2\Y^h\M^{-1} \right]\Biggl\} \, , \\
\label{eq:Bgg}
\tilde{A}_{\gamma\gamma}^{\rm BSM} \simeq
v\,\text{Im}\Biggl\{ \text{Tr}\left[ Q_R^2\M^{-1}\Y^h\right]
+ \text{Tr}\left[ Q_L^2\Y^h\M^{-1} \right]\Biggl\} \, , \\
\label{eq:AZg}
\begin{split}
A_{Z\gamma}^{\rm BSM} \simeq \frac{v}{c_w^2} \, \int_{0}^{+\infty}\!\!\!&\text{d}s\!\int_{0}^{1}\!\text{d}z\,2z^2\times \\
\text{Re}\Biggl\{ \,\text{Tr}&\left[ Q_R(T_R^3-s_w^2Q_R)e^{-sz\M^\dag\M} \M^\dag\Y^h e^{-s(1-z)\M^\dag\M} \right]\\
+\, \text{Tr}&\left[ Q_L(T_L^3-s_w^2Q_L)e^{-s(1-z)\M\M^\dag}\Y^h\M^\dag e^{-sz\M\M^\dag} \right]\Biggl\}
\end{split}\, , \\
\label{eq:BZg}
\begin{split}
\tilde{A}_{Z\gamma}^{\rm BSM} \simeq \frac{v}{c_w^2}\, \int_{0}^{+\infty}\!\!\!&\text{d}s\!\int_{0}^{1}\!\text{d}z\,2z~\times \\
\text{Im}\Biggl\{ \text{Tr}&\left[ Q_R(T_R^3-s_w^2Q_R)e^{-sz\M^\dag\M}\M^\dag\Y^he^{-s(1-z)\M^\dag\M} \right]\\
+ \text{Tr}&\left[ Q_L(T_L^3-s_w^2Q_L)e^{-s(1-z)\M\M^\dag}\Y^h\M^\dag e^{-sz\M\M^\dag} \right]\Biggl\} \ .
\end{split} \, 
\end{gather}
Here we simplified the expressions above making use of the relation
\be
\M Q_R - Q_L\M=\Y^h Q_R - Q_L\Y^h=0 \, ,
\ee
due to the fact that the electric charge correspond to an unbroken gauge symmetry and the Higgs field $h$ is uncharged under it.
Note that if the $h$ field couples to the fermions through the substitution $v\to v+h$, the Yukawa coupling matrix of the Higgs can be written as
\be
\Y^{h}=\frac{\partial}{\partial v}\M \, .
\ee

\subsection{Oblique corrections}

\begin{figure}[ht]
\centering
\includegraphics[width=6.5cm]{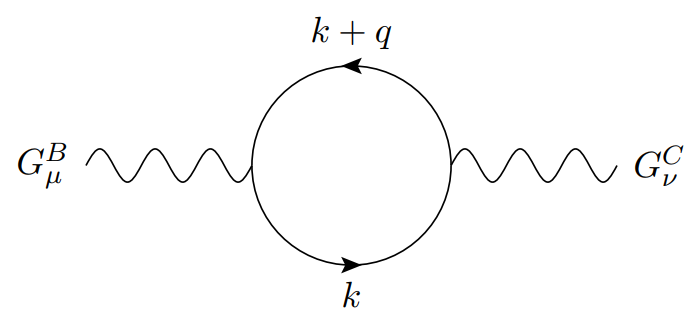}
\caption{Feynman diagram relative to the vacuum polarization functions in \eq{eq:PiBC}. }
\label{fig:dia2vacuum}       
\end{figure}

The gauge boson propagators get radiative correction from loops. Fermion contributions to the vacuum polarization functions come from 1-loop such as shown in \fig{fig:dia2vacuum}.
After integrating out the heavy fermion fields, one gets 
\beq\label{eq:PiBC}
\begin{split}
\Pi_{BC}^{\mu\nu}=&\sum_{\substack{i,j\\ \chi_{1},\chi_{2}}}\!g_{B}g_{C}(U_{\chi_{2}}^{\dag}Q_{\chi_{2}}^{B}U_{\chi_{2}})_{ji}(U_{\chi_{3}}^{\dag}Q_{\chi_{3}}^{C}U_{\chi_{3}})_{ij}\\
&\times i\mu^{4-d}\!\int\!\frac{\text{d}^{d}\!k}{(2\pi)^{d}}\,\frac{\text{Tr}_{\text{D}}[\gamma^{\mu}\!P_{\chi_{1}}\!(\slashed{k}+m_{i})\gamma^{\nu}\!P_{\chi_{2}}\!(\slashed{k}+\slashed{q}+m_{j})]}{[k^{2}-m_{i}^{2}]\,[(k+q)^{2}-m_{j}^{2}]} \\
=& \Pi_{BC}(q^2)g^{\mu\nu} + q^\mu q^\nu\text{-terms} \, ,
\end{split}
\eeq
where we introduced dimensional regularization. Expanding with $d=4-\epsilon$, one gets
\be
\begin{split}
\Pi_{BC}(q^2)=\Pi_{CB}(q^2)=\frac{g_Bg_C}{4\pi^2}\frac{1}{\epsilon}\Biggl\{
\frac{1}{2}\,&\text{Tr}\left[Q_R^B \M^\dag\M Q_R^C + Q_R^B  Q_R^C \M^\dag\M \right]\\
+\frac{1}{2}\,&\text{Tr}\left[Q_L^B \M\M^\dag Q_L^C + Q_L^B  Q_L^C \M\M^\dag \right]\\
-&\text{Tr}\left[Q_R^B \M^\dag Q_L^C \M + Q_L^B  \M Q_R^C \M^\dag \right]\\
-\frac{q^2}{3}&\text{Tr}\left[Q_R^B Q_R^C  + Q_L^B  Q_L^C \right]\Biggl\}\\
-\frac{g_Bg_C}{8\pi^2}\int_0^1\!\text{d}\!z\,\int_{0}^{+\infty}\!\!\!\text{d}s\,\frac{1}{s} \Biggl\{
2z(1-z)q^2e^{sz(1-z)q^2}\,&\text{Tr}\left[Q_R^B e^{-s(1-z)\M^\dag\M} Q_R^C e^{-sz\M^\dag\M}\right]\\
+2z(1-z)q^2e^{sz(1-z)q^2}\,&\text{Tr}\left[ Q_L^B e^{-s(1-z)\M\M^\dag} Q_L^C e^{-sz\M\M^\dag} \right]\\
-2z(1-z)q^2e^{-s\tilde{\mu}^2}\,&\text{Tr}\left[Q_R^B Q_R^C  + Q_L^B  Q_L^C \right]\\
-ze^{sz(1-z)q^2}\,&\text{Tr}\left[Q_R^B e^{-s(1-z)\M^\dag\M} Q_R^C \M^\dag\M e^{-sz\M^\dag\M}\right]\\
-ze^{sz(1-z)q^2}\,&\text{Tr}\left[ Q_L^B e^{-s(1-z)\M\M^\dag} Q_L^C \M\M^\dag e^{-sz\M\M^\dag} \right]\\
+ze^{-s\tilde{\mu}^2}\,&\text{Tr}\left[Q_R^B Q_R^C\M^\dag\M  + Q_L^B  Q_L^C\M\M^\dag \right]\\
-(1-z)e^{sz(1-z)q^2}\,&\text{Tr}\left[Q_R^B\M^\dag\M e^{-s(1-z)\M^\dag\M} Q_R^C  e^{-sz\M^\dag\M}\right]\\
-(1-z)e^{sz(1-z)q^2}\,&\text{Tr}\left[ Q_L^B\M\M^\dag e^{-s(1-z)\M\M^\dag} Q_L^C  e^{-sz\M\M^\dag} \right]\\
+(1-z)e^{-s\tilde{\mu}^2}\,&\text{Tr}\left[Q_R^B \M^\dag\M Q_R^C  + Q_L^B  \M\M^\dag Q_L^C \right]\\
+e^{sz(1-z)q^2}\,&\text{Tr}\left[Q_R^B e^{-s(1-z)\M^\dag\M}\M^\dag Q_L^C \M  e^{-sz\M^\dag\M}\right]\\
+e^{sz(1-z)q^2}\,&\text{Tr}\left[ Q_L^B e^{-s(1-z)\M\M^\dag}\M Q_R^C \M^\dag  e^{-sz\M\M^\dag} \right]\\
-e^{-s\tilde{\mu}^2}\,&\text{Tr}\left[Q_R^B \M^\dag Q_L^C\M  + Q_L^B  \M Q_R^C \M^\dag \right]\Biggl\} \, ,
\end{split}
\ee
where $\tilde{\mu}=4\pi e^{-\gamma_E}\mu$, 
with $\gamma_E$ being the Euler–Mascheroni constant.

\subsubsection{Peskin-Takeuchi parameters}

The Peskin-Takeuchi parameters $S$ and $T$ are defined through the vacuum polarization functions $\Pi(q^2)$ as
\begin{align}
T \equiv & \frac{1}{\alpha} \left( \frac{\Pi_{WW}^{\text{BSM}}(0)}{m_W^2} - \frac{\Pi_{ZZ}^{\text{BSM}}(0)}{m_Z^2} \right) \, , \\
S\equiv & \frac{4s_w^2c_w^2}{\alpha}\left[ \frac{\Pi_{ZZ}^{\text{BSM}}(m_Z^2)-\Pi_{ZZ}^{\text{BSM}}(0)}{m_Z^2} - \frac{c_w^2-s_w^2}{c_ws_w}\frac{\Pi_{Z\gamma}^{\text{BSM}}(m_Z^2)}{m_Z^2} - \frac{\Pi_{\gamma\gamma}^{\text{BSM}}(m_Z^2)}{m_Z^2} \right] \nonumber \\
\simeq & \frac{4s_w^2c_w^2}{\alpha}\left[ \Pi_{ZZ}^{\prime,\text{BSM}}(0) - \frac{c_w^2-s_w^2}{c_ws_w}\Pi_{Z\gamma}^{\prime,\text{BSM}}(0) - \Pi_{\gamma\gamma}^{\prime,\text{BSM}}(0) \right] \, ,
\end{align}
where the last approximation holds if the new physics is much heavier than the SM bosons.
We then obtain
\be
\label{eq:Tgen}
\begin{split}
T=\frac{1}{4\pi s_w^2c_w^2m_Z^2}\int_0^1\!\text{d}\!z\,\int_{0}^{+\infty}\!\!\!\text{d}s\,\frac{1}{s} \Biggl\{
(1-z)\,&\text{Tr}\left[T_R^+e^{-sz\M^\dag\M}T_R^-e^{-s(1-z)\M^\dag\M}\M^\dag\M\right]\\
+(1-z)\,&\text{Tr}\left[T_R^-e^{-sz\M^\dag\M}T_R^+e^{-s(1-z)\M^\dag\M}\M^\dag\M\right]\\
-4(1-z)\,&\text{Tr}\left[T_R^3e^{-sz\M^\dag\M}T_R^3e^{-s(1-z)\M^\dag\M}\M^\dag\M\right]\\
+(1-z)\,&\text{Tr}\left[T_L^+e^{-sz\M\M^\dag}T_L^-e^{-s(1-z)\M\M^\dag}\M\M^\dag\right]\\
+(1-z)\,&\text{Tr}\left[T_L^-e^{-sz\M\M^\dag}T_L^+e^{-s(1-z)\M\M^\dag}\M\M^\dag\right]\\
-4(1-z)\,&\text{Tr}\left[T_L^3e^{-sz\M\M^\dag}T_L^3e^{-s(1-z)\M\M^\dag}\M\M^\dag\right]\\
+\,&\text{Tr}\left[T_R^+e^{-sz\M^\dag\M}\M^\dag T_L^- e^{-s(1-z)\M\M^\dag}\M\right]\\
+\,&\text{Tr}\left[T_R^-e^{-sz\M^\dag\M}\M^\dag T_L^+ e^{-s(1-z)\M\M^\dag}\M\right]\\
-4\,&\text{Tr}\left[T_R^3e^{-sz\M^\dag\M}\M^\dag T_L^3 e^{-s(1-z)\M\M^\dag}\M\right]\Biggl\} \ , 
\end{split}
\ee
where $T^{\pm}=(T^1\pm iT^2)/\sqrt{2}$ are the non-diagonal isospin generators,
and
\be
\label{eq:Sgen}
\begin{split}
S\simeq\frac{2}{\pi}\int_0^1\!\text{d}\!z\,z(1-z)\,\int_{0}^{+\infty}\!\!\!\text{d}s\, \Biggl\{\,&\text{Tr}\left[T_R^3e^{-sz\M^\dag\M}\M^\dag T_L^Y\M e^{-s(1-z)\M^\dag\M}\right]\\
+\,&\text{Tr}\left[T_L^3e^{-sz\M\M^\dag}\M T_R^Y \M^\dag e^{-s(1-z)\M\M^\dag}\right]\\
-(1-z)\,&\text{Tr}\left[T_R^3e^{-sz\M^\dag\M}T_R^Ye^{-s(1-z)\M^\dag\M}\M^\dag\M\right]\\
-(1-z)\,&\text{Tr}\left[T_R^Ye^{-sz\M^\dag\M}T_R^3e^{-s(1-z)\M^\dag\M}\M^\dag\M\right]\\
-(1-z)\,&\text{Tr}\left[T_L^3e^{-sz\M\M^\dag}T_L^Ye^{-s(1-z)\M\M^\dag}\M\M^\dag\right]\\
-(1-z)\,&\text{Tr}\left[T_L^Ye^{-sz\M\M^\dag}T_L^3e^{-s(1-z)\M\M^\dag}\M\M^\dag\right]\\
+\frac{1}{s}\,&\text{Tr}\left[T_R^3e^{-sz\M^\dag\M}T_R^Ye^{-s(1-z)\M^\dag\M}\right]\\
+\frac{1}{s}\,&\text{Tr}\left[T_L^3e^{-sz\M\M^\dag}T_L^Ye^{-s(1-z)\M\M^\dag}\right]\Biggl\} \, ,
\end{split}
\ee
where $T^Y$ is the hypercharge generator.

\begin{small}

\bibliographystyle{utphys}
\bibliography{bibliography.bib}

\end{small}

\end{document}